# Evolution at two levels of gene expression in yeast

Carlo G. Artieri and Hunter B. Fraser*

Department of Biology, Stanford University, Stanford, CA 94305, USA.

*Corresponding author

CONTACT INFORMATION:

Hunter B. Fraser
Herrin Labs Rm 305
371 Serra Mall
Stanford, CA 94305
United States

TELEPHONE NUMBER: 650-723-1849

FAX NUMBER: 650-724-4980

EMAIL: hbfraser@stanford.edu

RUNNING HEAD: Evolution at two levels of gene expression

NUMBER OF FIGURES: 4

NUMBER OF TABLES: 2

NUMBER OF SUPPLEMENTS: 1 Document, 4 Tables

**Abstract**

Despite the greater functional importance of protein levels, our knowledge of gene expression evolution is based almost entirely on studies of mRNA levels. In contrast, our understanding of how translational regulation evolves has lagged far behind. Here we have applied ribosome profiling—which measures both global mRNA levels and their translation rates—to two species of *Saccharomyces* yeast and their interspecific hybrid in order to assess the relative contributions of changes in mRNA abundance and translation to regulatory evolution. We report that both *cis* and *trans*-acting regulatory divergence in translation are abundant, affecting at least 35% of genes. The majority of translational divergence acts to buffer changes in mRNA abundance, suggesting a widespread role for stabilizing selection acting across regulatory levels. Nevertheless, we observe evidence of lineage-specific selection acting on a number of yeast functional modules, including instances of reinforcing selection acting at both levels of regulation. Finally, we also uncover multiple instances of stop-codon readthrough that are conserved between species. Our analysis reveals the underappreciated complexity of post-transcriptional regulatory divergence and indicates that partitioning the search for the locus of selection into the binary categories of 'coding' vs. 'regulatory' may overlook a significant source of selection, acting at multiple regulatory levels along the path from genotype to phenotype.

**Introduction**

Almost four decades ago it was argued that coding sequence changes were insufficient to explain the morphological divergence between humans and chimpanzees, suggesting that changes in gene expression regulation may have played a dominant role (King and Wilson 1975). More recently, a major focus of modern evolutionary genetics has been to understand the molecular basis of regulatory variation within and between species (Carroll 2005; Rockman and Kruglyak 2006). In almost all instances, however, 'regulatory variation' has been used synonymously with 'differences in mRNA levels'—despite decades of research indicating that post-transcriptional regulation is essential (Day and Tuite 1998). Developments in quantitative proteomics have uncovered patterns of divergence at the level of the proteome both within and between species, and a unifying observation has been that mRNA abundance is an imprecise proxy of protein abundance (e.g., de Souza Abreu et al. 2009; Khan et al. 2012; Wu et al. 2013; Skelly et al. 2013). Indeed, when the contributions of mRNA abundance were accounted for, these studies found that protein levels were independent, heritable phenotypes, confirming that regulatory evolution beyond the level of mRNA is common. Nevertheless, the relatively low power and high cost of these approaches have limited their use in dissecting the molecular bases of regulatory variation between closely related species. This has encouraged a focus on mRNA levels—aided by the availability of high-throughput transcriptional profiling methods (e.g. microarrays and RNA-seq)—which has left many fundamental questions about the evolution of translational dynamics unanswered.

It has long been known that natural selection generates synonymous codon usage bias (CUB) in favor of codons represented by the most abundant tRNAs (Ikemura 1981, Plotkin and Kudla 2011), perhaps to enhance the speed and/or accuracy of protein translation (Akashi 2003). Both intra- and interspecies comparative studies have found that purifying selection appears to be the dominant mode of evolution acting at the level of CUB (Drummond and Wilke 2008, Zhou et al. 2010, Waldman et al. 2011). Nevertheless, potentially adaptive changes have been observed, such as an increase in CUB among cytosolic ribosomal proteins and glycolytic enzymes in anaerobic yeasts, coinciding with their shift to primarily fermentative growth (Man and Pilpel 2007). While these studies highlight the action of natural selection beyond the transcriptional level, the effect of CUB on translation is still actively debated (Tuller et al. 2011; Ingolia et al. 2011; Qian et al. 2012; Charneski and Hurst 2013), making the biological significance of such findings difficult to interpret.

Encouragingly, a wealth of insight about translational regulation has surfaced via the application of a new method enabling the measurement of protein translation rates of the coding transcriptome (Ingolia et al. 2009; Ingolia 2010). Termed 'ribosome profiling' (or riboprofiling), it involves isolating and sequencing short fragments of mRNA bound by actively translating ribosomes and provides quantitative information about the translational states of all transcripts. Riboprofiling has revealed that relative translational rates vary across the transcriptome by approximately 100-fold in the budding yeast *Saccharomyces cerevisiae*, contributing substantially to the dynamic range of expression (Ingolia et al. 2009). Furthermore, the translation of individual genes can be modulated in response to external conditions such as nutrient starvation or meiosis (Ingolia et al. 2009; Brar et al. 2012). Therefore abundant opportunity exists for regulatory variation in translational efficiency; however, how this evolves within and between species remains unknown.

Both transcriptional and translational regulation can diverge via changes in *cis*-regulatory elements (CREs), or through changes affecting the *trans*-acting regulatory factors that bind these elements. The relative contributions of each mechanism to divergence can be dissected via measurement of individual allelic expression levels in interspecific hybrids (Wittkopp 2005; Muller and Nieduszynski 2012). The common *trans* environment shared by the two alleles in hybrids means that any differences in allele-specific expression (ASE) must reflect changes in CREs. The fraction of expression divergence not attributable to ASE is therefore the result of changes in *trans*-acting factors (in the absence of epistatic *cis* x *trans* interactions). Though much more is known about transcriptional CREs (see Wittkopp and Kalay 2011), similar *cis*-acting mechanisms regulate the rate of translation (Gebauer and Hentze 2004). A recent study measuring protein levels in yeast hybrids using mass-spectrometry found both *cis*- and *trans*-acting effects, but with divergence detected at fewer than 100 genes (Khan et al. 2012), it is difficult to extend these conclusions to the whole transcriptome.

Because changes in CRE activities can be highly temporally and spatially specific—in contrast to amino acid changes that typically alter a protein everywhere it is expressed—it has been suggested that regulatory adaptation may primarily occur through changes in *cis*-regulation (Carroll 2005; Lemos et al. 2008). However, identifying those regulatory changes that have occurred due to the action of selection has traditionally proven to be challenging (Fraser 2011; Barrett and Hoekstra 2011). Several studies have applied methods to detect accelerated expression divergence in large-scale datasets (e.g., Rifkin et al. 2003, Gilad et al. 2006), however detecting selection has not been possible in the absence of an accurate null model of neutral divergence in gene expression. More recently, a novel approach to identifying instances of selection on gene expression was introduced, and takes advantage of the observation that most phenotypes are polygenic—resulting from the action of multiple functionally related genes (Weiss 2008). Significant bias in the directionality of ASE in a hybrid (favoring one parent's alleles) among multiple members of a functionally related group of genes indicates that multiple coordinated *cis*-acting mutations have occurred and is evidence of selection acting in a lineage-specific manner (Fraser et al. 2010; Bullard et al. 2010; Fraser 2011). Analysis of ASE in hybrids has been used to identify hundreds of genes subject to lineage-specific selection, including several complexes and pathways in domesticated yeasts (Fraser et al. 2010; Bullard et al. 2010), pathogenic adaptations in clinical yeasts (Fraser et al. 2012), as well as morphological, physiological, and behavioral adaptations between strains of mice (Fraser et al. 2011).

Here we apply a similar framework to study the impact of natural selection on translation using ribosome profiling in hybrids of closely related species of yeast. Unlike previous studies of translational divergence, which have either used codon usage as a proxy for translational efficiency (e.g., Man and Pilpel 2007), or have had limited statistical power and/or coverage of the proteome (Khan et al. 2012), this approach captures ribosomal occupancy directly, and therefore takes into account the potential for changes in the rate of initiation or pausing. Furthermore, as ribosome profiling generates ASE information for both mRNA abundance and translational efficiency simultaneously (Ingolia et al. 2009), it offers an unparalleled opportunity to compare patterns of divergence across both levels, thereby offering a glimpse into the landscape of regulatory divergence beyond mRNA abundance.

**Results**

*Simultaneous detection of regulatory divergence at two levels*

In order to compare *cis*-regulatory divergence in yeasts at the levels of mRNA abundance and translation simultaneously, we performed ribosome profiling (Ingolia et al. 2009, Ingolia 2010) on the interspecific hybrid of *S. cerevisiae* and its closely related wild congener, *S. paradoxus* (~5 million years diverged) (Scannell et al. 2011). Ribosome profiling involves the construction of two RNA-seq libraries from each sample: the first is derived from poly-adenylated mRNA (hereafter called the 'mRNA' fraction) and measures the abundance of each mRNA in the cell. The second library is derived from fragments of these mRNAs protected from nuclease digestion by actively translating ribosomes (the 'Ribo' fraction). As more highly transcribed genes produce more read counts in both the mRNA and Ribo fractions, the relative translational efficiency (hereafter simply 'translation') of each coding mRNA is determined by dividing its abundance in the Ribo fraction by its corresponding abundance in the mRNA fraction (both measured in reads per kilobase per million mapped reads, or RPKM). Ratios greater than one indicate transcripts with higher than average translation (per mRNA transcript) while ratios lower than one reflect transcripts with lower levels of translation (Ingolia et al. 2009).

After performing ribosome profiling for two biological replicates in nutrient-rich conditions (see Methods), we mapped reads to a set of 4,640 high-confidence 1:1 orthologs (Scannell et al. 2011) for which most reads could be unambiguously assigned to one of the parental alleles (see Methods; Supplemental Table S1). As expected, Ribo fractions showed an overwhelming preference for the protein-coding regions of mRNAs, and biological replicate abundance measurements and estimated translational efficiency from both fractions agreed well (Spearman's $\rho > 0.97$ and $\rho > 0.85$ for estimated abundances and translational efficiencies, respectively; Supplemental Table S2, Supplemental Fig. S1). Furthermore, the distributions of RPKMs in both fractions are not significantly different between species, indicating that there is no systematic bias in allelic abundances favoring either species (Kruskal-Wallis rank sum test, p = 0.99 and 0.88 for the mRNA and Ribo fractions, respectively).

Within hybrids, both alleles share the same *trans*-acting cellular environment. Therefore, ASE in the mRNA fraction is indicative of *cis*-regulatory divergence of mRNA abundance between species (denoted as hybrid $^{Sc}/_{Sp}$ mRNA ≠ 1, where $^{Sc}/_{Sp}$ indicates the ratio of the *S. cerevisiae* allele's expression level to that of *S. paradoxus*) (Fig 1A). Similarly, the translational *cis* ratio refers to the ratio between the Ribo ASE and the mRNA ASE. In the absence of *cis*-regulatory divergence in translational efficiency, the ASE ratio of the Ribo fraction should equal that of the mRNA fraction. Therefore, significant *cis*-regulatory divergence in translation is inferred when these ratios differ (hybrid $^{Sc}/_{Sp}$ Ribo ≠ hybrid $^{Sc}/_{Sp}$ mRNA). As our inference of translational divergence includes variability in the estimates of $^{Sc}/_{Sp}$ ratio from both fractions, it likely has reduced power to detect significant differences relative to mRNA abundance (see below).

Furthermore, estimates of Ribo ASE may be less accurate than mRNA ASE, because of both lower read counts (Supplemental Table S2) and greater heterogeneity within transcripts, likely due to variation in ribosomal processivity (Ingolia et al. 2009). Indeed, estimates of hybrid $^{Sc}/_{Sp}$ were more reproducible between biological replicates in the mRNA fraction (Spearman's ρ = 0.78 and 0.58, for the mRNA and Ribo fractions, respectively; Supplemental Fig. S2). Therefore we applied a previously developed test of *cis*-regulatory divergence to the mRNA level (Supplemental Fig. S3; Bullard et al. 2010) and modified it to account for this difference at the translational level (see Methods). Briefly, in order to detect significant translational *cis*-regulatory divergence (i.e., hybrid $^{Sc}/_{Sp}$ Ribo ≠ hybrid $^{Sc}/_{Sp}$ mRNA), we applied a resampling approach that takes into account differences between alleles in base composition, length, and read coverage (Fig. 1B). This approach was more conservative than simply testing for significant differences from binomial expectations of read coverage (Supplemental Fig. S4).

*Cis-regulatory divergence in translation is pervasive*

In order to compare patterns of regulatory divergence between mRNA abundance and translation directly, we restricted our analysis to the 3,665 orthologs to which at least 100 reads mapped across both alleles in both fractions (see Methods). Our estimates of ASE in mRNA abundance agreed with a previous microarray-based analysis of this hybrid (Tirosh et al. 2009; Supplemental Fig. S5). Significant *cis*-regulatory divergence in translational efficiency was detected in 35% of orthologs, as compared to 61% with significant divergence in mRNA abundance (Fig. 2A, Supplemental Fig. S6). However this apparently greater role of divergence in mRNA abundance is largely a result of our conservative approach to detecting translational divergence, leading to greater statistical power to detect divergence at this level. When comparing the magnitudes of divergence in mRNA abundance vs. translation, we actually find a slightly stronger role for translation (median absolute $\log_2$ *cis* ratio = 0.325 for translation and 0.288 for mRNA abundance; Kruskal-Wallis $p = 0.009$). This suggests that translation efficiency may be of comparable importance as mRNA abundance in the evolution of protein production rates in yeast.

Among those orthologs with significant *cis*-regulatory divergence in both mRNA abundance and translation, changes at the two levels could either be reinforcing (acting in the same direction) or opposing (acting in opposite directions). For neutral changes not influenced by natural selection, an equal number of each would be expected (Fraser et al 2010). However we found a greater than two-fold excess of genes whose divergence is in opposing directions at the two regulatory levels (561 opposing vs. 256 reinforcing, $\chi^2$ test $p = 7.1 \times 10^{-27}$), leading to maintenance of similar protein abundances between species (Fig. 2A). We found no evidence that this was biased by extreme measurements, as both reinforcing and opposing divergence were observed across the full range of expression levels and *cis* ratios (Supplemental Material; Supplemental Fig. S7). In order to address this phenomenon more generally, we compared the $^{Sc}/_{Sp}$ ratios calculated from both fractions across all orthologs and found that changes in mRNA abundance tend to overestimate the divergence in protein production rate by ~15% (Fig. 2B).

Interestingly, comparison with a dataset of mRNA expression variability across 17 *S. cerevisiae* strains grown under nutrient-rich conditions (Kvitek et al. 2008) revealed that orthologs with opposing *cis*-acting divergence were significantly less variable than orthologs with reinforcing differences (Kruskal-Wallis rank sum test, p = 0.030; p = 0.0072 for strongly opposing changes, defined as the 50% of genes with the largest differences in Ribo-mRNA *cis* ratios). Therefore, orthologs with opposing directionality of changes, in which translational differences tend to buffer mRNA level changes, are associated with genes that show more constrained mRNA abundances across strains of *S. cerevisiae*, consistent with the action of stabilizing selection. We also explored patterns of sequence divergence in the promoters, 5′ UTRs, CDSs, and 3′ UTRs among orthologs with reinforcing vs. opposing *cis*-acting divergence and found no significant differences between categories with the exception of slightly increased conservation of the 5′ UTRs of opposing orthologs (Kruskal-Wallis test, $p = 0.010$) (Supplemental Fig. S8).

A positive relationship has been reported between upstream sequence divergence and *cis*-acting divergence in mRNA levels, as expected if divergence in promoter elements underlies regulatory divergence (Tirosh et al. 2009). Controlling for confounding effects of divergence within the CDS, we found similar positive relationships between the absolute $^{Sc}/_{Sp}$ mRNA and translational *cis* ratios and sequence divergence in 5′ UTRs, with a slightly stronger effect in the latter ($p = 0.0042$ and $0.00069$ for the mRNA and translational levels, respectively; see Supplemental Material).

Previous studies have noted that promoters containing a TATA box—a key CRE that affects transcription initiation—tend to have greater divergence in mRNA levels than TATA-less promoters (Tirosh et al 2006; Landry et al. 2007; Tirosh et al. 2009; Skelly et al 2013). A similar effect has been found for promoters with high nucleosome occupancy proximal to their transcriptional start site (hereafter 'occupied proximal-nucleosome' or OPN; Tirosh et al. 2008, 2010; Tirosh and Barkai 2011). We observed independent positive relationships between TATA or OPN promoters and divergence at the mRNA level, but not in translation, suggesting that their effects are restricted to the level of transcription (Kruskal-Wallis rank sum test, $p = 0.00079$ and $0.00040$, for TATA and OPN at the mRNA level, and $p = 0.43$ and $0.82$ at the translational level, respectively; Fig. 2C). These results remained unchanged when considering both factors simultaneously (Supplemental Fig. S9). We also found a slight yet significant excess of TATA-less promoters among those with opposing as compared to reinforcing divergence at both regulatory levels (89% vs. 80% TATA-less for opposing and reinforcing divergence, respectively; $\chi^2$ test $p = 0.0026$), supporting the notion that these genes may be subject to stabilizing selection to preserve protein levels.

At the translational level, it has been noted that ribosomal occupancy is a function of the rate of ribosomal processivity, which differs across codons (Letzring et al. 2010). Highly expressed transcripts show strong codon usage bias (CUB), which has been hypothesized to ensure high translational efficiency (preventing sequestration of ribosomes) and/or accuracy (preventing the production of non-functional proteins; Gingold and Pilpel 2011). When controlling for mRNA level, a significant negative correlation was observed between CUB as measured in *S. cerevisiae* and the absolute translational *cis* ratio, but not the absolute mRNA *cis* ratio (analysis of covariance [ANCOVA], $p = 1.5 \times 10^{-12}$ and 0.13 respectively). The presence of mRNA secondary structure in the vicinity of the start codon has also been implicated in reducing translational efficiency (Kudla et al. 2009; Robbins-Pianka et al. 2010; Shah et al. 2013; Bentele et al. 2013; Dvir et al. 2013; Goodman et al. 2013). We found evidence for a positive correlation between species-specific decreases in computationally-predicted secondary structure in downstream of the start codon and increased translational efficiency (see Supplemental Material; Supplemental Fig. S10). Finally, we also note that several studies have suggested that the presence of translated upstream open reading frames (uORFs) in the 5′ UTRs of genes may regulate translational efficiency (Ingolia et al. 2009; Brar et al. 2012; Pelechano et al. 2013), however, we find no evidence that they play a significant role in explaining *cis*-regulatory divergence in translation between these species (Supplemental Material).

*Trans-acting regulatory divergence is widespread at both regulatory levels*

In the absence of epistasis between *cis* and *trans* regulation, the fraction of expression divergence not explained by *cis* divergence can be attributed to differences in *trans* acting factors (Wittkopp et al. 2004). In order to estimate the contribution of *trans* divergence at both regulatory levels, we performed riboprofiling on two biological replicates of the parental strains used to generate the hybrid and estimated the ratio of the *S. cerevisiae* ortholog's expression level to that of *S. paradoxus* (denoted as parental $^{Sc}/_{Sp}$ mRNA or Ribo; see Methods). As in the case of the hybrids, we observed high concordance between replicate measurements (Supplemental Figs. S11, S12; Supplemental Table S2). Following the same logic as above, the $^{Sc}/_{Sp}$ *trans* mRNA ratio is obtained by subtracting the log$_2$(hybrid $^{Sc}/_{Sp}$ mRNA) from log$_2$(parental $^{Sc}/_{Sp}$ mRNA). At the translational level, the *trans* ratio is obtained by subtracting the sum of the log$_2$ transformed hybrid Ribo ratio and the parental mRNA ratio from the interspecific difference in the Ribo fraction (parental $^{Sc}/_{Sp}$ Ribo - hybrid $^{Sc}/_{Sp}$ Ribo - parental $^{Sc}/_{Sp}$ mRNA), thereby accounting for mRNA differences between species as well as the fraction of translational divergence attributable to *cis* effects. Significant *trans* divergence at both levels was determined using the same resampling approach as above (see Methods).

To compare divergence in *cis* and *trans* across regulatory levels directly, we restricted the following analyses to the 3,634 orthologs with sufficient coverage in all samples and replicates (Supplemental Fig. S13; see Methods). Similar numbers of *cis* and *trans*-acting changes were detected for both mRNA (2,217 *cis* vs 2,384 *trans*) and Ribo (1,264 *cis* vs 1,275 *trans*).

Similar to our analysis of *cis*-acting divergence across regulatory levels (Fig. 2A), we tested for reinforcing or opposing patterns in *trans*. As was the case for the *cis* level, there was also a significant excess of opposing *trans* divergence across levels ($\chi^2$ test p = 5.1 × 10$^{-12}$; Supplemental Fig. S14). In addition, the patterns of *trans* divergence support the mRNA-level specific role of TATA boxes and OPNs (Supplemental Fig. S15), similar to our findings for *cis*-acting divergence (Fig. 2C). Supporting a general pattern of opposing mRNA and translational divergence that buffer changes in protein production rates, we found that the parental $^{Sc}/_{Sp}$ mRNA levels also overestimated the translational component of between-species regulatory divergence (Supplemental Fig. S16).

The reinforcing vs. opposing distinction can also be made comparing *cis* and *trans* divergence within a single regulatory level. As has been observed in this hybrid previously (Tirosh et al. 2009), there was a slight excess of opposing *cis* and *trans* changes at the mRNA level ($\chi^2$ test p = 0.0018; Supplemental Fig. S17). No significant difference was observed between reinforcing vs. opposing mechanisms of divergence at the translational level ($\chi^2$ test p = 0.83); however, this may reflect the lower precision of the Ribo fraction (see Supplemental Fig. S18).

*Polygenic selection at two levels of gene regulation*

We next determined whether there was evidence of lineage-specific polygenic selection in either mRNA abundance or translation by taking advantage of a recently developed approach to detect non-neutral evolution across functionally related groups of genes (Fraser et al. 2010, Bullard et al. 2010, Fraser 2011). Under neutral divergence of *cis*-regulation, no consistent bias is expected in the relative parental direction of ASE among genes within a functional category (e.g. a protein complex, biochemical pathway, or genes contributing to the same phenotype) (Fig. 3A). Conversely, consistent directional bias across a functional group indicates that multiple independent *cis*-regulatory changes have altered gene expression in a coordinated fashion, and is evidence of lineage-specific selection.

Therefore, we performed scans for selection independently at the level of mRNA abundance and translation, as well as among all orthologs with reinforcing directionality of bias at both regulatory levels. We tested 591 gene sets for deviation from neutral expected frequencies by means of a $\chi^2$ test, and employed a permutation framework to control for the number of tests performed (see Methods).

We detected lineage-specific enrichment in a number of functional categories representing a wide variety of cellular processes. In Table 2 we report the thirteen most significant gene sets (~1 expected by chance; full gene lists in Supplemental Table S3). Functions such as mating and telomeric silencing were found to be under lineage-specific selection on mRNA abundance, while for translation a protein complex involved in rRNA metabolism was implicated. Combining both levels, we found several gene sets with evidence for reinforcing lineage-specific selection on both mRNA abundance and translation, including kinases and genes related to heavy metal sensitivity (Table 2). Our finding of natural selection on both levels of regulation, in some cases targeting the same gene sets, highlights the importance of considering both levels simultaneously.

We then sought to determine whether any of the candidate instances of polygenic selection detected above was associated with phenotypic differences between these strains. One of the functional categories biased towards *S. cerevisiae*, 'divalent cations and heavy metals sensitivity' (Fig. 3B), harbors many genes involved in vacuolar regulation and transport. Since deleting these genes leads to deficient growth in the presence of divalent cations and heavy metals, we predicted that the *S. cerevisiae* lineage would exhibit increased resistance to these metals.

A recent study of yeast growth rates across 200 different conditions included the parental strains we used to generate the hybrid (Warringer et al. 2011). Among these were two different concentrations (denoted here as 'low' and 'high') of four divalent heavy metal cations: cadmium ($CdCl_2$), cobalt ($CoCl_2$), copper ($CuCl_2$), and nickel ($NiCl_2$). As predicted, *S. cerevisiae* strain S288c outperformed *S. paradoxus* CBS432 under all concentrations and metabolites in terms of growth rate, with the exception of nickel, where the difference between strains was negligible (Fig. 3C). In fact, the relative growth advantage of *S. cerevisiae* in high concentrations of copper and both concentrations of cobalt are among the largest phenotypic differences found between these strains (Warringer et al. 2011). Interestingly, the superior resistance to heavy metals of the *S. cerevisiae* parental strain does not appear to be a fixed difference between species, since many wild *S. cerevisiae* strains are less fit than their *S. paradoxus* counterparts in the presence of these cations (Warringer et al. 2011). Therefore the reinforcing *cis*-regulatory divergence observed across regulatory levels may reflect selection acting in a strain-specific manner, rather than species-level divergence.

*Identification of conserved C-terminal peptide extensions*

Organisms have been shown to increase peptide diversity by infrequent stop codon readthrough, one form of which involves the ribosome inserting an amino acid into the growing peptide at a stop codon position and continuing in-frame translation (von der Haar and Tuite 2006). Consequences of readthrough include prevention of deadenylation increasing mRNA stability, ribosome stalling inducing mRNA degradation, or production of a protein with a C-terminal peptide extension. Two functional C-terminal extensions were previously identified in *S. cerevisiae*: Extension of the *PDE2* gene decreases its stability, resulting in accumulation of cyclic AMP (Namy et al. 2002); and readthrough of *IMP3*, involved in ribosomal biogenesis, destabilizes its interaction with the U3 snoRNA (Cosnier et al. 2011). A recent systematic study of conserved protein-coding potential in candidate C-terminal extensions in eukaryotes failed to identify any candidates in yeasts (Jungreis et al. 2011), however the authors required strong sequence conservation of the extension across five *sensu stricto* species. Multi-species riboprofiling data provide an excellent opportunity to search for direct evidence of translation in putative C-terminal extensions at the transcriptome-wide level.

We identified all orthologs in which both species shared the potential for C-terminal peptide extensions via the presence of in-frame stop codons in their 3′ UTRs and assessed these putative C-terminal extensions for the presence of translation in the Ribo fractions (see Methods). Translation was detected in one or both species in 109 and 81 cases, respectively. The putative C-terminal extensions for all 190 genes were aligned and evaluated for their potential to encode conserved peptides by the absence of frame-shifting indels and CDS divergence patterns consistent with purifying selection.

These criteria identified 19 strong candidates for conserved C-terminal peptide extensions, representing a wide variety of functions including glycolysis (*PGK1*), response to heat shock (*AHA1*), actin filament stabilization (*TPM2*), and the large ribosomal subunit (four genes, hypergeometric test of enrichment $p = 2.2 \times 10^{-6}$) (Table 2, Supplemental Table S4). Interestingly, we detected *IMP3* readthrough only in *S. cerevisiae*, and the peptide sequence of the extension is not conserved (see Discussion). Translation was not detected in the C-terminal extension of *PDE2* in either species; however it is in the bottom quartile of translational efficiency among yeast genes, making detection of its estimated ~2.2% frequency of readthrough (Namy et al. 2002) challenging without very deep read coverage.

An example of conserved C-terminal extension, translation initiation factor eIF1A (*TIF11*), is shown in Fig. 4A. Tif11 is an essential protein that is involved in start codon identification whose C-terminus interacts with Fun12, a GTPase also involved in initiation of translation. Stop codon readthrough could potentially play a role in the regulation of this interaction. A number of species-specific readthrough events were also observed (Fig. 4B), suggesting this may be an unappreciated source of regulatory divergence.

**Discussion**

*Evolution at two regulatory levels*

A complete understanding of the role of regulatory change in the evolution of phenotypic diversity requires approaches to measuring divergence beyond the mRNA level. Using ribosome profiling of interspecific hybrids, we have identified *cis*- and *trans*-regulatory changes at two regulatory levels simultaneously. In particular, our results suggest that *cis*-acting divergence at the translational level is a common yet underappreciated feature of regulatory evolution. Indeed, despite our observation of a larger proportion of orthologs with significant divergence at the mRNA level (Fig. 2A), the magnitudes of the *cis*-ratios were similar at both levels, indicating that we have likely underestimated the frequency of translational divergence. This is supported by recent studies that have identified quantitative trait loci associated with protein abundance (pQTLs; Wu et al. 2013, Skelly et al. 2013), which have found that only approximately half of pQTLs can be explained by differences in transcript abundance, suggesting a substantial role for post-transcriptional regulation.

In cases where divergence occurred at both regulatory levels, we observed a dominant pattern of opposing directionality of change (both in *cis* and in *trans*), indicating that mRNA levels tend to overestimate the regulatory divergence in protein abundance in hybrids and the total divergence between species (Figs. 2A,B; Supplemental Fig. S16). Furthermore, this phenomenon was associated with genes that show constrained mRNA abundances across strains of *S. cerevisiae* (Kvitek et al. 2008), consistent with the action of stabilizing selection. Previous studies of mRNA abundance have established that stabilizing selection is the primary mode of selection acting upon the transcriptome (Rifkin et al. 2005; Denver et al. 2005; Bedford and Hartl 2009). The target of selection is likely protein abundance rather than mRNA expression level *per se*, and our results suggest that regulatory output may be canalized via changes at multiple levels.

Previous studies have found functional associations between divergence patterns in different regulatory mechanisms. For instance, Dori-Bachash et al. (2011) noted that divergence of transcription and mRNA degradation are often coupled, and controlled by the same regulators. In contrast, our findings indicate that control of mRNA levels and translation can result from different underlying architectures (e.g. related to TATA boxes and promoter nucleosomes). Interestingly a recent analysis of mRNA and protein abundance across 22 strains of *S. cerevisiae* found that the presence of TATA boxes was associated with greater inter-strain variability in both transcript and protein levels, the latter measured by tandem mass spectrometry (Skelly et al. 2013). Our results are consistent with this, and suggest that the relationship between TATA promoters and divergence in protein abundance results from their effect at the transcriptional level (Fig. 2C), in line with the well-established role of the TATA box (Tirosh et al. 2006, 2008, 2009).

Similarly, other factors may act only at the translational level. For example, translational rate is thought to vary along individual transcripts due to codon translation rate variability and/or mRNA secondary structures (Kertesz et al. 2010; Gingold and Pilpel 2011), in contrast to the more nearly constant rate of transcriptional elongation (Singh and Padgett 2009). Though analysis of ribosomal profiling data from multiple species has produced equivocal results regarding the effect of codons on translational elongation rates (Tuller et al. 2011; Ingolia et al. 2011; Qian et al. 2012; Charneski and Hurst 2013), we found an association between high CUB and conservation of translational efficiency, providing evolutionary evidence that codon usage is associated with translational dynamics. Furthermore, we observed that *cis*-acting translational differences are associated with changes in computationally predicted secondary structure (Supplemental Fig. S10).

*Polygenic selection at multiple regulatory levels*

Our observation of lineage-specific ASE bias across functional groups provides the first direct evidence of polygenic selection on translation, and indicates that such selection can be reinforcing across multiple regulatory levels. Similar to the McDonald-Kreitman test (McDonald and Kreitman 1991), our test will detect any lineage-specific difference in selection pressure, so an open question is which of these cases represent positive selection, as opposed to a relaxation of negative selection in one lineage. Although signatures of recent selective sweeps have been used to infer adaptation in similar cases when comparing strains within *S. cerevisiae* (Fraser et al 2010, Fraser et al 2012), this approach has little power for the far more ancient divergence of the lineages we have studied here.

However regardless of the mode of lineage-specific selection at work, these regulatory changes may have led to divergence in diverse phenotypes. The gene set with the clearest phenotypic connection—higher levels of both mRNA and translation in *S. cerevisiae* among genes whose loss leads to heavy metal sensitivity—makes the prediction that *S. cerevisiae* may have greater tolerance to these metals, which is indeed the case (Fig. 3C). As noted above, this tolerance to heavy metals is not a fixed difference between the species, but rather is specific to some domesticated strains of *S. cerevisiae*. In particular, the superior tolerance of domesticated strains to growth in high copper environments has long been thought to reflect artificial selection imposed by brewing in copper containers as well as the use of copper sulfate as a fungicide and insecticide (Fogel and Welch 1982). Although the amplification of the *CUP1* gene is a major source of this resistance (Warringer et al. 2011), many genes are involved in metal tolerance – some unique to specific cations and others shared by multiple (Bleackley et al. 2011) – and our results suggest that the ancestors of S288c may have experienced a history of polygenic adaptation for this trait.

Another notable example of lineage-specific selection involves the mating/fertilization gene set, in which 20 genes have higher mRNA abundance from *S. paradoxus* alleles, compared to only three from *S. cerevisiae* alleles. Interestingly, while sexual reproduction is thought to be rare in the wild for both species, estimates of mating frequency are ~50-fold higher for *S. paradoxus* as compared to *S. cerevisiae* (Tsai et al. 2008; Ruderfer et al. 2006), consistent with either selection to increase expression in *S. paradoxus*, or perhaps relaxed constraint on their *cis*-regulation in *S. cerevisiae*. However for the majority of gene sets with evidence of lineage-specific selection (Table 1), we could not make any specific phenotypic predictions.

*Conservation and divergence of C-terminal peptide extensions*

C-terminal peptide extensions via stop codon readthrough are thought to play a relatively minor role in eukaryotic proteomic diversity, as only a handful of experimentally observed examples are known (Jungreis et al. 2011). Combining direct translational evidence from the Ribo fraction with sequence conservation between the parental species, we identified 19 candidates for conserved C-terminal extensions (Table 2). However, in the majority of cases where translation was detected in putative extensions, the peptide sequence was poorly conserved (62 cases) and/or species-specific (109 cases; Supplementary Table 4), including the experimentally verified extension of *IMP3* (Cosnier et al. 2011).

Our observations suggest two features of C-terminal extensions in yeasts: First, conserved peptide extensions may not require sequence conservation to be functional. Both verified extensions in yeast exert their effects by destabilizing protein function and/or interactions (Namy et al. 2002; Cosnier et al. 2011). This may result from the addition of any unstructured component to the C-terminus, which can lead to destabilization and degradation of the folded polypeptide (von der Haar and Tuite 2006). Second, it has been suggested that peptide extensions represent a mechanism for organisms to transiently expose hidden genetic information (True et al. 2004). If functional, C-terminal peptide extensions may evolve rapidly because of their ability to be transiently expressed in response to specific conditions, employing a translational mechanism to mitigate the potentially deleterious costs of changes in the constitutively translated portion of the peptide.

*Towards a comprehensive view of gene expression evolution*

Although we have discovered widespread natural selection contributing to the divergence of translation rates, complementing the extensive literature on the evolution of mRNA abundances, these two levels still represent only a fraction of the steps from DNA to protein. Other regulatory mechanisms such as mRNA splicing/editing/localization/decay, post-translational modification, and protein decay, are all likely targets of natural selection as well. As technologies able to probe these levels continue to be developed, a more holistic understanding of how gene expression evolves will be achievable. We speculate that transcription and translation (together with alternative splicing in some species) may emerge as the dominant levels at which selection shapes protein abundances, owing to the exquisite spatial and temporal specificity achievable by minor alterations of the multitude of discrete *cis*-regulatory elements controlling these two regulatory levels.

**Materials and Methods**

*Yeast strains and growth conditions*

A diploid interspecific hybrid yeast strain was produced by mating the haploid strains of *S. cerevisiae* (isogenic to BY4716 MATα *lys2* ura::KAN) and *S. paradoxus* CBS432 (MATa ura::HYG). All samples and replicates were derived from single-colonies grown in YPD medium at 30°C. Two biological replicates of the hybrid and parental strains were collected during log phase growth ($OD_{600}$ 0.5 - 0.7) from 750 ml YPD cultures grown in a C24 Incubator Shaker (New Brunswick Scientific) at 30ºC for at least 16 hours.

*Ribosome profiling library construction and sequencing*

Ribosome profiling next-generation sequencing (NGS) libraries were prepared as detailed in Ingolia (2010) with modifications by Brar et al. (2012) and sequenced to a read length of 36 bases using an Illumina HiSeq 2000 instrument at the Stanford Center for Genomics and Personalized Medicine (see Supplemental Materials). All data are deposited in the NCBI Sequence Read Archive under accession #SRP028614.

*Allele-specific read mapping*

Hybrid and parental reads from both fractions were mapped using Bowtie version 0.12 (Langmead et al. 2009) in a strand-specific manner using the iterative method described in Ingolia (2010) in order to enrich for ribosome protected fragments and account for spurious adenine (A) bases added to the 3′ ends of reads by the oligo-dT mediated reverse transcription (see Supplemental methods). All analyses of coverage were restricted to locations where all possible reads spanning the nucleotide of interest would map uniquely. Furthermore, we removed all nucleotides within 27 bp of a splice junction as junction-spanning reads were likely to be underrepresented in our short read lengths.

Mapping reads were assigned to genomic locations (CDS, 5′ and 3′ UTRs, or introns) based on the position of their 5′ most base. Criteria follow Ingolia et al. (2009), except for the CDS (16 bases upstream of the first nucleotide and 16 bases upstream of the last nucleotide) and 3′ UTRs (13 bases upstream of the first nucleotide to 15 bases upstream of the last nucleotide) in order to minimize the possibility that reads assigned to the latter were spurious signal from the CDS during our analysis of candidate stop codon readthrough (see below).

*Genome assemblies and annotation*

*S. cerevisiae* and *S. paradoxus* genome assemblies, annotations, and orthology assignments were obtained from (Scannell et al. 2011), from which we curated a high-confidence set of 4,640 nuclear genes (Supplemental Material; Supplementary Table S1).

*Detecting significant cis-regulatory divergence in hybrids*

We first obtained base-level read coverages in the CDSs of both species for all uniquely mappable positions for all hybrid fractions and replicates for the 4,640 orthologs. A minimum of 100 reads mapping among both alleles within each replicate mRNA fraction (4,436 orthologs) or each replicate in both the mRNA and Ribo fractions (3,665 orthologs) were required to test for evidence of mRNA and translational *cis* regulatory divergence, respectively. To test for significant *cis* differences in the mRNA abundance ($^{Sc}/_{Sp}$ mRNA ≠ 1), we implemented the resampling test detailed in Bullard et al. (2010; Supplemental Fig. S2; Supplemental Material). For the test of significant *cis* regulatory divergence in translation (as shown in Fig. 1B), we sought to reject the null hypothesis that $\log_2(^{Sc}/_{Sp}$ Ribo) was not significantly different from $\log_2(^{Sc}/_{Sp}$ mRNA). Therefore, we resampled the CDS base-level coverage of the *S. cerevisiae* allele using the *S. cerevisiae* marginal nucleotide frequencies ($\pi_c = \pi_c[A], \pi_c[C], \pi_c[G], \pi_c[T]$) and length ($L_c$) and the *S. paradoxus* allele using $\pi_p$ and $L_p$ 10,000 times in the each replicate of the Ribo fraction. Each resampling was used to generate a distribution of started $\log_2$ ratios (total base level coverage from $\pi_c, L_c + 1$ / total base level coverage from $\pi_p, L_p + 1$), denoted as $\log_2(^{Sc+1}/_{Sp+1})$, which takes into account the variability in read coverage across each allele. These distributions were then compared to the observed $\log_2(^{Sc+1}/_{Sp+1}$ mRNA) ratio in the same replicate to generate a p-value based on how often the observed ratio was outside the bounds of the permuted distribution. The same resampling was then repeated reciprocally in each mRNA fraction replicate, which was compared to observed $\log_2(^{Sc+1}/_{Sp+1}$ Ribo) ratio in the same replicate. If the directionality of difference agreed among all comparisons, the least significant of the four p-values was retained. Finally, p-values were adjusted such that we retained only those comparisons significant at an FDR of 5% for further analysis (Benjamini and Hochberg 1995). We employed an equivalent approach to detect *trans*-divergence using the parental data (Supplemental Material).

*Analysis of factors associated with* cis-*regulatory divergence*

We obtained Table S3 from Kvitek et al. (2008) and calculated the corrected coefficient of covariation ($[1+1/4n] \times$ COV) across the mean-centered expression coefficients for the 17 strains analyzed. The distributions of corrected COV were then compared among orthologs with significant reinforcing or opposing divergence at both regulatory levels. *S. cer* and *S. par* promoter (the 200 nt upstream of the TSS), 5′ UTR, CDS, and 3′ UTR sequences were aligned using DIALIGN-TX version 1.0.0 (Subramanian et al. 2005) and pairwise % divergence (1 - % identity) was calculated according to method four of Raghava and Barton (2006), which considers only internal but not terminal gaps. For correlations, pairwise tests, and multiple regressions, only orthologs with sufficient numbers of mapping reads to be tested for significance were analyzed. The multiple regression model was lm(|$^{Sc}/_{Sp}$ mRNA or translational *cis*| ~ Promoter %DIV + 5′ UTR %DIV+ 3′ UTR %DIV + CDS %DIV), where '% DIV' stands for % divergence. The presence or absence of a TATA box or OPN in the promoter was determined for each gene in our dataset represented in Tirosh et al. (2006 and 2008) and used to test for an association with increased absolute *cis* ratio using the Kruskal-Wallis rank sum test. In order to analyze the effects of TATA boxes and OPNs individually, we analyzed genes containing either one or the other element, but not both, independently (as shown in Fig. 2C; Supplemental Fig. S9 shows the same analysis when not excluding orthologs that have both elements). As a measure of CUB, we obtained the codon bias index (CBI) values from the SGD for each ortholog with an SGD identifier. Because CBI is associated with mean mRNA fraction RPKM across alleles and replicates (Spearman's $\rho = 0.615$, $p < 10^{-15}$), the relationship between absolute divergence in *cis* ratio was determined by analysis of covariance, including mean mRNA fraction RPKM as a covariate.

*Detecting lineage-specific cis-regulatory divergence*

Orthologs with significant *cis*-regulatory divergence at either level were divided into two categories based on the upregulating parental allele and ranked based on the magnitude of their absolute *cis* ratio (from largest to smallest). In order to increase our power to detect selection among genes with reinforcing bias, we used the replicate averaged mRNA and translational *cis* ratios to identify reinforcing divergence among all orthologs that passed our threshold for analysis at both regulatory levels (e.g., 3,665). Any replicates whose direction of reinforcement differed between replicates (< 2%) were removed. Reinforcing orthologs were ranked as above using the sum of their $\log_2$ *cis* ratios. This resulted in three ranked gene sets consisting of *S. cerevisiae* and *S. paradoxus* biased orthologs: mRNA abundance, translation, and reinforcing.

We searched for lineage-specific bias among the following functional categories represented in the FunSpec database (Robinson et al. 2002): Gene Ontology (GO) biological process, GO molecular function, GO cellular component, MIPS functional category, MIPS phenotypes, MIPS complexes, MIPS protein classes, and PFAM domains. In order to detect lineage-specific bias within a gene set, we identified all functional categories containing at least 10 members in the set and determined whether significant bias existed in the direction of one or the other lineage using a $\chi^2$ 'goodness of fit' test. Because many different categories were being tested, we determined the probability of observing a particular enrichment by permuting ortholog assignments and repeating the test 10,000 times, retaining the most significant p-value observed in each functional dataset. A category specific FDR was obtained by asking how often a p-value of equal or greater significance would be observed in the permuted data. The test of bias was performed on three difference thresholds, using either the top 25 or 50% most biased orthologs along each parental lineage, or analyzing all biased orthologs. In the case where a functional category was shared in two datasets and the test was performed on the exact same orthologs, only the category with the lowest FDR was reported.

The analysis of the data of Warringer et al. (2011) was performed on the growth rate measurement in Dataset S1. The *S. cerevisiae* BY4716 strain used in this study is isogenic to strain S288C, which was used for the comparison to *S. paradoxus* strain CBS432.

*Identification of candidate 3′ UTR C-terminal extensions*

The *S. cerevisiae* 3′ UTRs identified by Nagalakshmi et al. (2008) and sequence of equivalent length downstream of the stop codon of *S. paradoxus* orthologs were scanned for an in-frame stop codon (TAA, TAG, or TGA) at least 5 codons downstream of the canonical stop in both species. Orthologs with downstream in-frame stop codons in both species were retained for analysis (but see below). Because of the low number of reads mapping to 3′ UTRs, we applied a number of different criteria to identify instances of readthrough (Supplemental Materials; Supplemental Table S4 lists all potential C-terminal extensions that at least meet the criteria for single-species translation). In the case of *PDE2*, a gene previously identified to experience functional readthrough (Namy et al. 2002) we identified a frameshift indel that extends the C-terminal extension to 32 amino acids in *S. paradoxus* as compared to 22 in *S. cerevisiae*. This required extending 81 bp annotated 3′ UTR to at least 96 bp in *S. paradoxus*.

*Statistics*

All statistics were performed using R version 2.14.0 (R Core Team 2013). Kruskal-Wallis rank sum tests were performed using 10,000 permutations of the data as implemented in the 'coin' package (Hothorn et al. 2008). FDRs for significant *cis* regulatory divergence were calculated using the Benjaminin and Hochberg method (1995) using the p.adjust() command.

**Data Access**

All raw sequencing reads are deposited in the NCBI Gene Expression Omnibus under accession #GSE50049.

**Acknowledgements**

We are indebted to the members of Patrick O. Brown's lab at Stanford, in particular Dustin Hite and Dan Klass, for sharing resources, equipment, and advice. We also thank members of the Fraser and Petrov labs in addition to three anonymous reviewers for useful comments on earlier versions of this manuscript. This work was supported by an NSERC Postdoctoral Fellowship to CGA and NIH grant 1R01GM097171-01A1. HBF is a Sloan Fellow and Pew Scholar.

**Author Contributions**

CGA and HBF designed the study. CGA performed all experimental work and data analysis. CGA and HBF wrote the manuscript.

**Disclosure Declaration**

The authors declare no conflicts of interest.

**Figure Legends**

**Figure 1.** (A) Identifying *cis*-regulatory divergence at two levels. In the example, the *S. paradoxus* allele (blue) is transcribed at a higher level than that of *S. cerevisiae* (red), as represented by the larger number of wavy lines. However, the *S. cerevisiae* allele has higher translational efficiency, as represented by the larger number of ribosomes per transcript (pairs of grey circles). The *S. paradoxus* mRNA *cis* bias manifests as a negative $\log_2(^{Sc}/_{Sp})$ ratio in the mRNA fraction. If translational efficiency was unchanged between alleles, the more abundant allele, in this case *S. paradoxus*, would produce more footprints in the Ribo fraction. Therefore the translational *cis* ratio is obtained by dividing the $^{Sc}/_{Sp}$ Ribo fraction ratio by the mRNA fraction ratio (which is equivalent to a subtraction in $\log_2$). The net $\log_2(^{Sc}/_{Sp})$ translational *cis* ratio is positive, indicating *cis* bias favoring *S. cerevisiae* translation. (B) Detection of significant translational divergence is based upon rejecting the null hypothesis that the observed allelic ratios are not significantly different from one another (see A). The observed $^{Sc}/_{Sp}$ ratios (red circles, mRNA fraction; blue circles, Ribo fraction) (i) were obtained directly from the replicates of the two fractions. (ii) These were permuted by resampling the base-level coverage of each allele with replacement 10,000 times, generating a distribution of $^{Sc}/_{Sp}$ ratios that captures the inter-allelic variability in base-composition, length, and read coverage. (iii) The distributions of permuted ratios (boxplots) were then each reciprocally compared to the corresponding observed ratio (e.g., the permuted distribution of $^{Sc}/_{Sp}$ Ribo ratios [blue boxplots] was compared to the observed mRNA $^{Sc}/_{Sp}$ ratio [red circles] and vice-versa) for which a two-tailed p-value was calculated. If all comparisons agreed in the parental direction of allelic bias, then (iv) the least significant p-value (indicated by the red asterisk) was used as the representative for the comparison. See Supplemental Material for application of the test to the mRNA level and *trans* comparisons.

**Figure 2.** (A) The relationship between *cis*-regulatory divergence at the mRNA abundance and translational levels (all plotted $^{Sc}/_{Sp}$ ratios are the mean of the two biological replicates). Divergence was detected only at the mRNA level for a large fraction of genes (orange circles), though greater than one tenth of orthologs were significantly diverged only in translation (blue circles). Among orthologs diverged at both levels, we observed a significant excess opposing (red triangles) as compared to reinforcing changes (green squares). The number of orthologs in each class is indicated in the barplot. *S. cer*, *S. cerevisiae*; *S. par*, *S. paradoxus*. (B) Opposing divergence across regulatory levels. The red line indicates the best fit of a linear regression, with equation, p, and $r^2$ values indicated above. The slope is significantly lower than one (95% confidence interval ±0.033), indicating that $^{Sc}/_{Sp}$ mRNA ratio estimates tend to overestimate the degree of difference by ~15% relative to that of the Ribo fraction. (C) Orthologs whose promoters contain either TATA-boxes (TATA) or occupied proximal nucleosome regions (OPN; Tirosh et al. 2008) show more divergence in *cis* only at the mRNA level when compared to non-TATA promoters (Non) or depleted proximal nucleosome regions (DPNs), respectively. Kruskal-Wallis test p-values are indicated above each fraction.

**Figure 3.** (A) Detecting selection from patterns of ASE in hybrids. The example above shows ASE levels (indicated by the wavy lines) for four genes belonging to a particular functional category. Black "X"s indicate downregulating *cis*-regulatory differences between the parental alleles. For a given group of functionally related genes evolving neutrally, no bias is expected with respect to the directionality of ASE in hybrids (No selection). However, biased directionality, as in the case where all down-regulating mutations occurred along the *S. cerevisiae* lineage (Selection), indicates a history of lineage-specific selection acting on *cis*-regulation. (B) Reinforcing lineage-specific bias on orthologs involved in divalent cation and heavy metal resistance. Green triangles indicate orthologs within this functional category with reinforcing directionality of bias at both regulatory levels. Significantly more (17) orthologs are reinforcing along the *S. cerevisiae* lineage as compared to that of *S. paradoxus* (5). All orthologs are indicated as grey circles. (C) *S. cerevisiae* strain S288c is more resistant to heavy metals than *S. paradoxus* strain CBS432. Shown are the $\log_2$ transformed relative growth rates (*S. cerevisiae*/*S. paradoxus*) for the four heavy metals at two concentrations (L, low; H, high) measured by Warringer et al. (2011). *S. cerevisiae* outperforms *S. paradoxus* under all conditions, though in the presence of nickel, the difference is negligible.

**Figure 4.** Evidence of stop codon readthrough leading to C-terminal peptide extension. The translation initiation codons are indicated by the right-facing arrow, the annotated ORF by the thick black lines, and the canonical stop codon by the black triangles. The candidate C-terminal peptide extension is indicated by the grey line terminated by in-frame stop codons in the 3′ UTR (grey triangles above the line for *S. cerevisiae*, and below for *S. paradoxus*). Dark shades (red, *S. cerevisiae*; blue, *S. paradoxus*) indicate nucleotide-level coverage of mRNA fraction reads and light shades indicate Ribo fraction reads. (A) Example of conserved C-terminal peptide extension of the translation initiation factor eIF1A (*TIF11*). The putative 21 amino acid extension is conserved and well covered by reads in the Ribo fraction of both species. (B) Example of a *S. paradoxus* specific C-terminal extension in *MRPS16*, a subunit of the mitochondrial ribosome. mRNA fraction reads indicate that the 3′ UTR is expressed in both species; however, translation is only detected in the 17 amino acid extension of *S. paradoxus*, and not the potential 21 amino acid extension of *S. cerevisiae*. Interestingly, coverage of the C-terminal extension in *S. paradoxus* is comparable to that of the CDS, suggesting that readthrough of this gene may be frequent.

## Tables

**Table 1.** Functional categories with evidence of polygenic selection on gene regulation. Tests were performed independently at the mRNA abundance or translational levels, or among orthologs with reinforcing directionality of bias at both levels. Three thresholds (top 25%, 50%, or all orthologs) based on the magnitude of the $\log_2(^{Sc}/_{Sp})$ ratio were tested. Group, specific identifier for category within dataset; Direction, parent with the most upregulating alleles; Sc | Sp, the number of upregulating alleles in *S. cerevisiae* | *S. paradoxus*; FDR, the probability that a category would be observed by chance as determined from 10,000 permutations of the data (see Methods). The top thirteen most significant categories are shown. Abbreviations of functional datasets are as follows: MIPS Func. Cats, Munich Information Center for Protein Sequences Functional Categories; Prot. Classes, Protein Classes. Approximately 1 false positive is expected by chance based on summation of the FDRs.

| Functional Dataset | Group | Annotation | Direction | Sc \| Sp | FDR |
|---|---|---|---|---|---|
| **mRNA *cis*** | | | | | |
| TOP 25% | | | | | |
| GO Component | GO:0031225 | anchored to membrane | *S. cer* | 9 \| 2 | 0.16 |
| ALL | | | | | |
| GO Process | GO:0006348 | chromatin silencing at telomere | *S. cer* | 24 \| 8 | 0.15 |
| GO Function | GO:0004175 | endopeptidase activity | *S. cer* | 10 \| 1 | 0.13 |
| MIPS Func. Cats. | 41.01.01 | mating \|fertilization\| | *S. par* | 4 \| 22 | 0.04 |
| MIPS Complexes | 550.2.132 | Unknown | *S. cer* | 17 \| 4 | 0.09 |
| PFAM Domains | SH3_1 | Cytoskeletal regulation | *S. cer* | 11 \| 1 | 0.07 |
| | AAA | Mitochondrial Rho GTPases | *S. cer* | 14 \| 2 | 0.05 |
| **Translational *cis*** | | | | | |
| ALL | | | | | |
| GO Function | GO:0016740 | transferase activity | *S. cer* | 87 \| 57 | 0.13 |
| MIPS Complexes | 550.2.140 | Ribosomal RNA metabolism | *S. par* | 2 \| 15 | 0.09 |
| **Reinforcing divergence at both regulatory levels** | | | | | |
| TOP 50% | | | | | |
| GO Function | GO:0016301 | kinase activity | *S. par* | 3 \| 18 | 0.03 |
| GO Process | GO:0006260 | DNA replication | *S. par* | 1 \| 12 | 0.12 |
| MIPS Complexes | 550.1.108 | Protein synthesis/turnover | *S. par* | 2 \| 10 | 0.05 |
| ALL | | | | | |
| MIPS Phenotype | 62.35.02 | Divalent cations and heavy metals sensitivity | *S. cer* | 17 \| 5 | 0.19 |

**Table 2.** List of candidate orthologs with conserved C-terminal peptide extensions.

| SGD ID | Name | Details |
| --- | --- | --- |
| YBR025C | OLA1 | P-loop ATPase with similarity to human OLA1 and bacterial YchF, identified as specifically interacting with the proteasome |
| YBR283C | SSH1 | Subunit of the Ssh1 translocon complex, Sec61p homolog involved in co-translational pathway of protein translocation |
| YCR012W | PGK1 | 3-phosphoglycerate kinase, catalyzes transfer of phosphoryl groups from 1,3-bisphosphoglycerate to ADP to produce ATP |
| YDR214W | AHA1 | Co-chaperone that binds to Hsp82p and activates its ATPase activity |
| YER056C-A | RPL34A | Ribosomal 60S subunit protein L34A |
| YGL031C | RPL24A | Ribosomal 60S subunit protein L24A |
| YIL138C | TPM2 | Minor isoform of tropomyosin, binds to and stabilizes actin cables and filaments |
| YJL158C | CIS3 | Mannose-containing glycoprotein constituent of the cell wall |
| YLR175W | CBF5 | Pseudouridine synthase catalytic subunit of box H/ACA small nucleolar ribonucleoprotein particles |
| YLR340W | RPP0 | Conserved ribosomal protein P0 of the ribosomal stalk |
| YLR390W-A | CCW14 | Covalently linked cell wall glycoprotein |
| YML028W | TSA1 | Thioredoxin peroxidase, acts as both a ribosome-associated and free cytoplasmic antioxidant |
| YMR260C | TIF11 | Translation initiation factor eIF1A |
| YMR307W | GAS1 | Beta-1,3-glucanosyltransferase, required for cell wall assembly and also has a role in transcriptional silencing |
| YOL086C | ADH1 | Alcohol dehydrogenase, fermentative isozyme active as homo- or heterotetramers |
| YOL143C | RIB4 | Lumazine synthase, catalyzes synthesis of immediate precursor to riboflavin |
| YPL061W | ALD6 | Cytosolic aldehyde dehydrogenase |
| YPL131W | RPL5 | Ribosomal 60S subunit protein L5 |
| YPL234C | VMA11 | Vacuolar ATPase V0 domain subunit c', involved in proton transport activity |

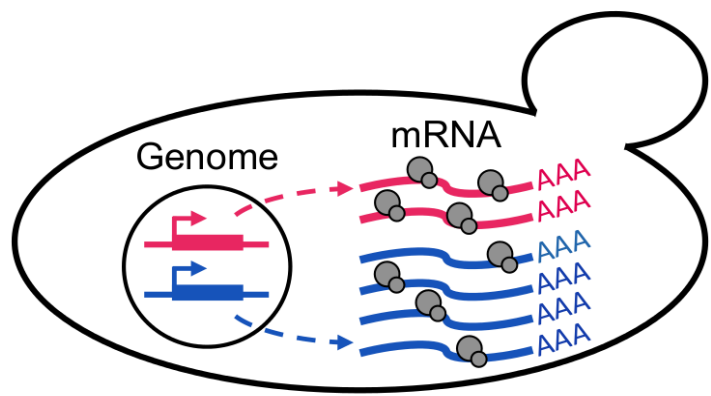
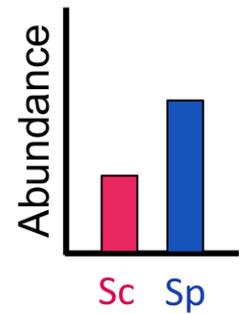
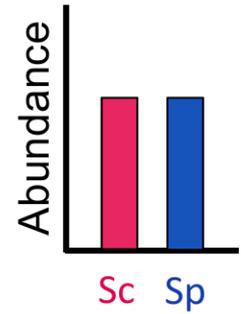
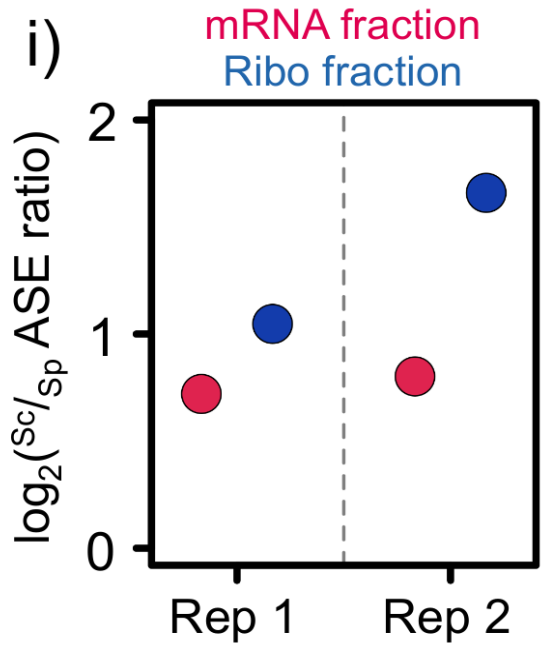
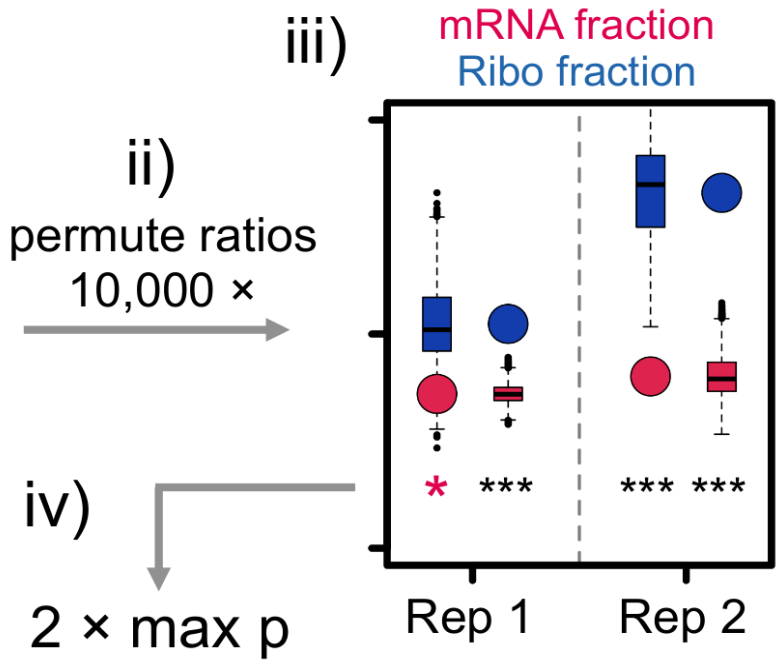

FIG 1

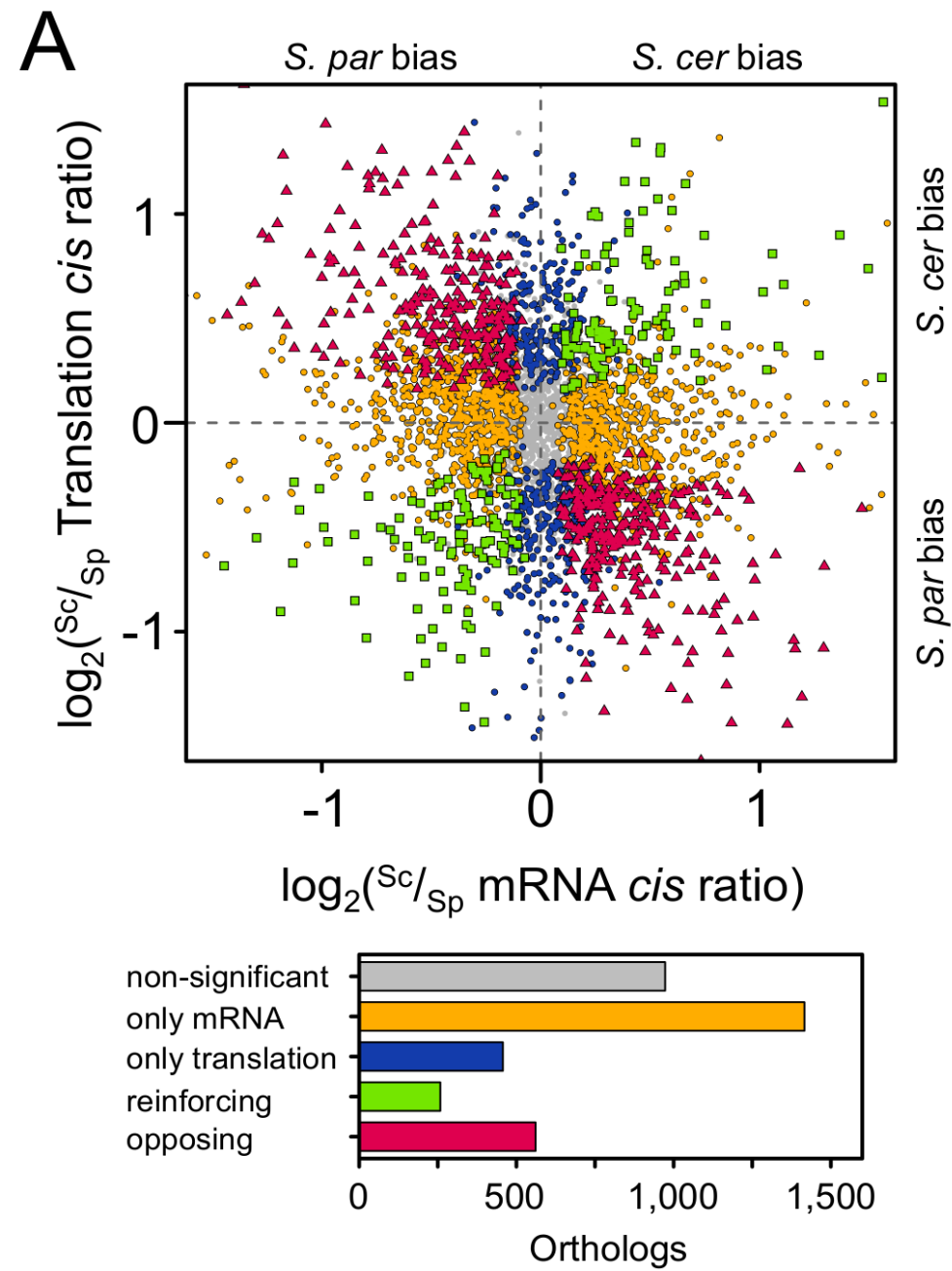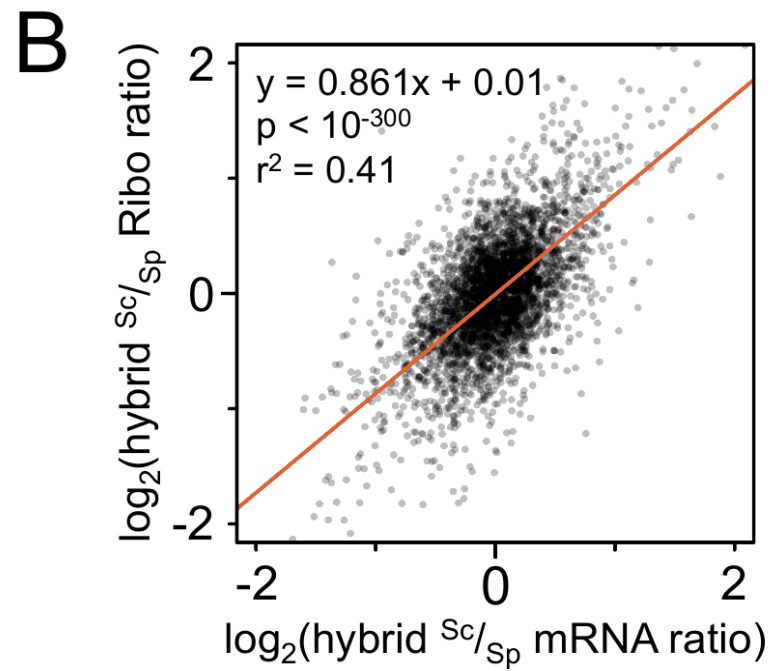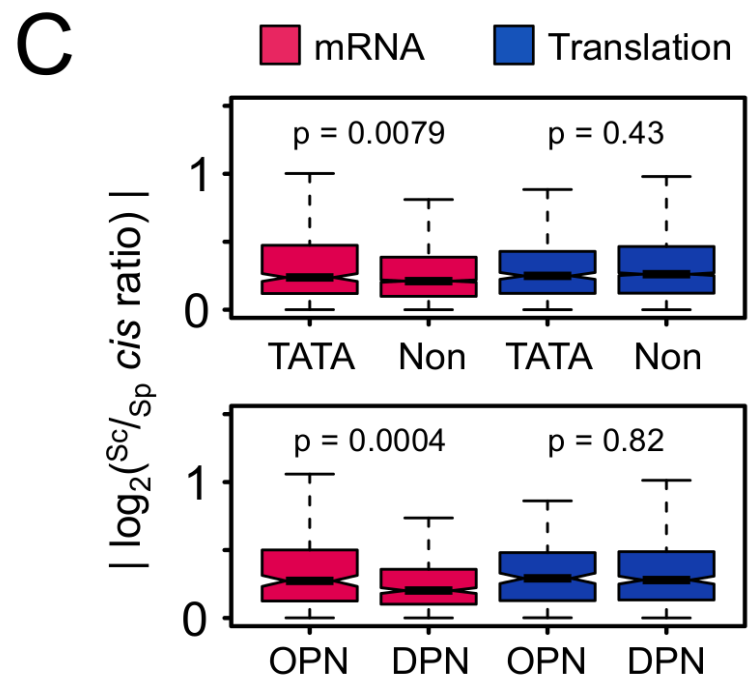

FIG 2

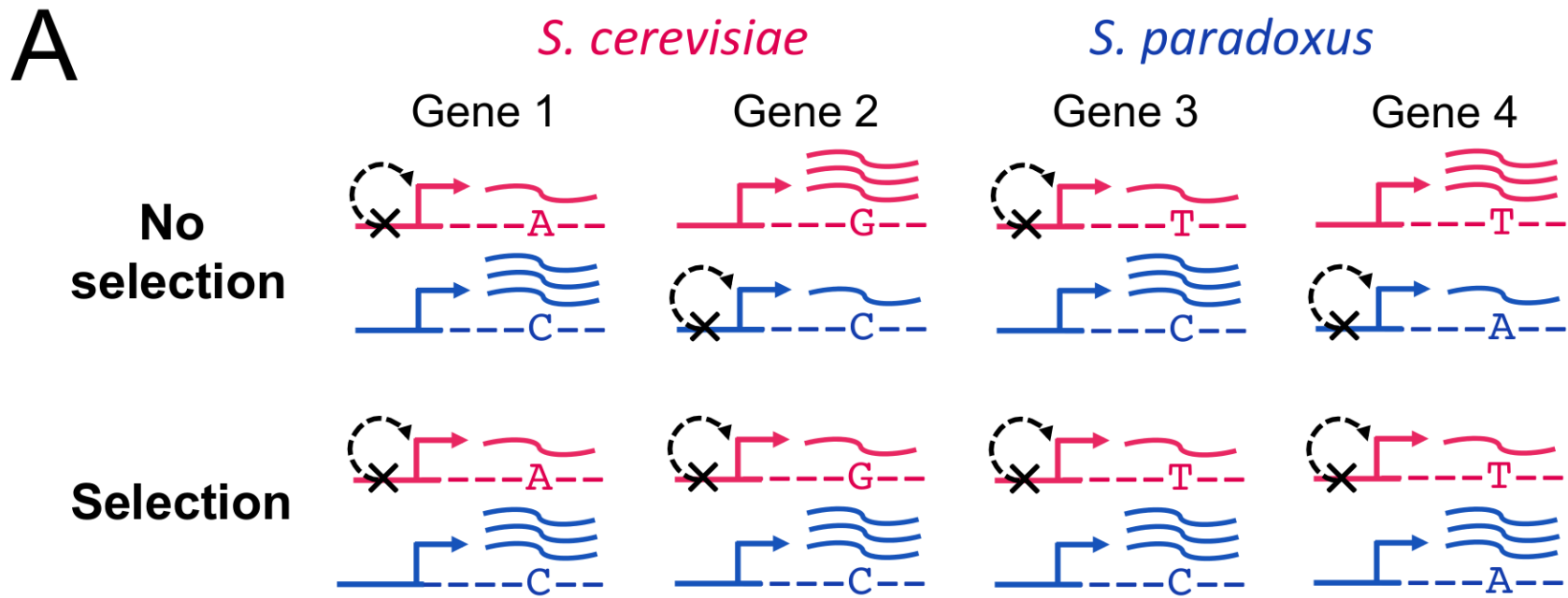
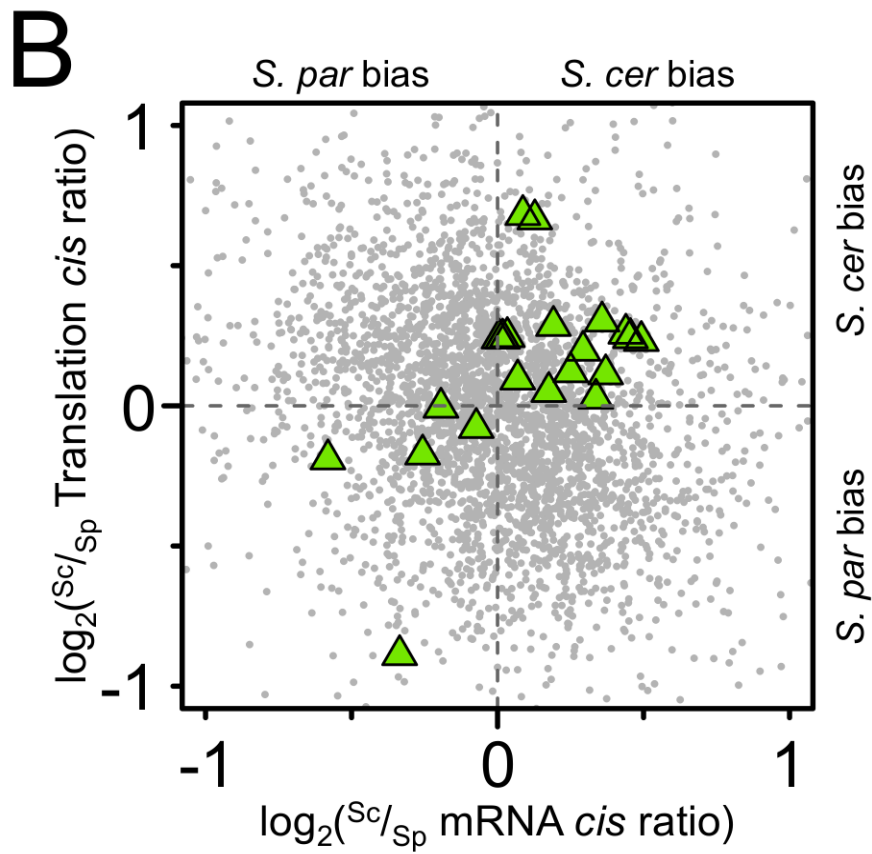
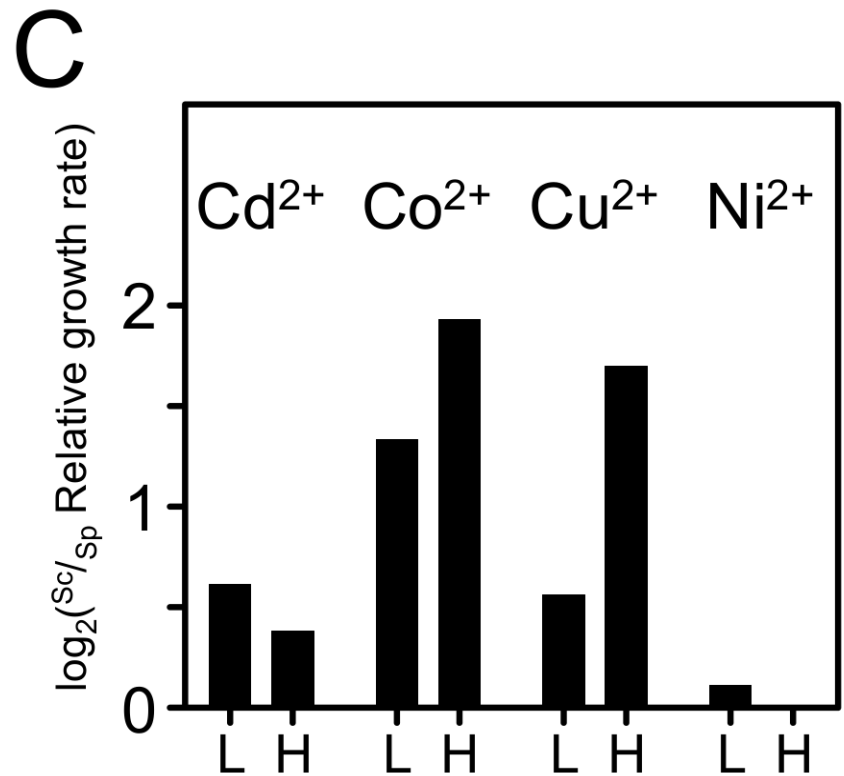

FIG 3

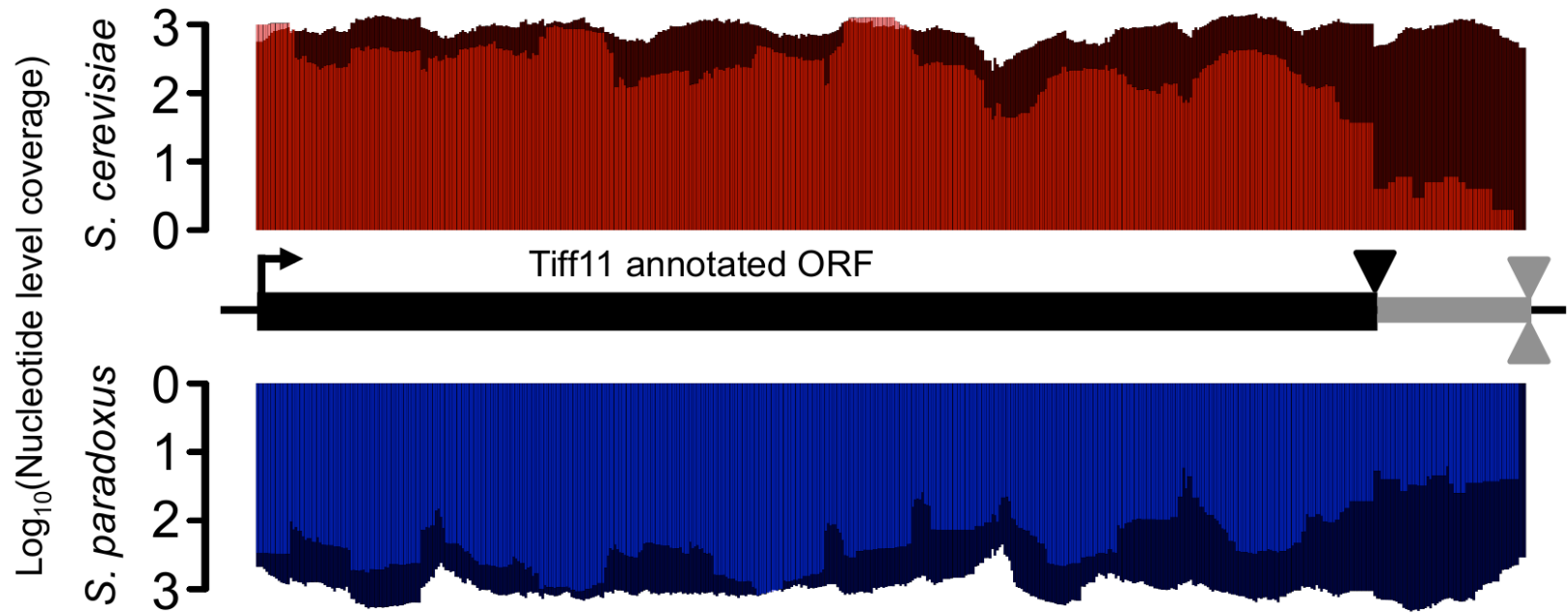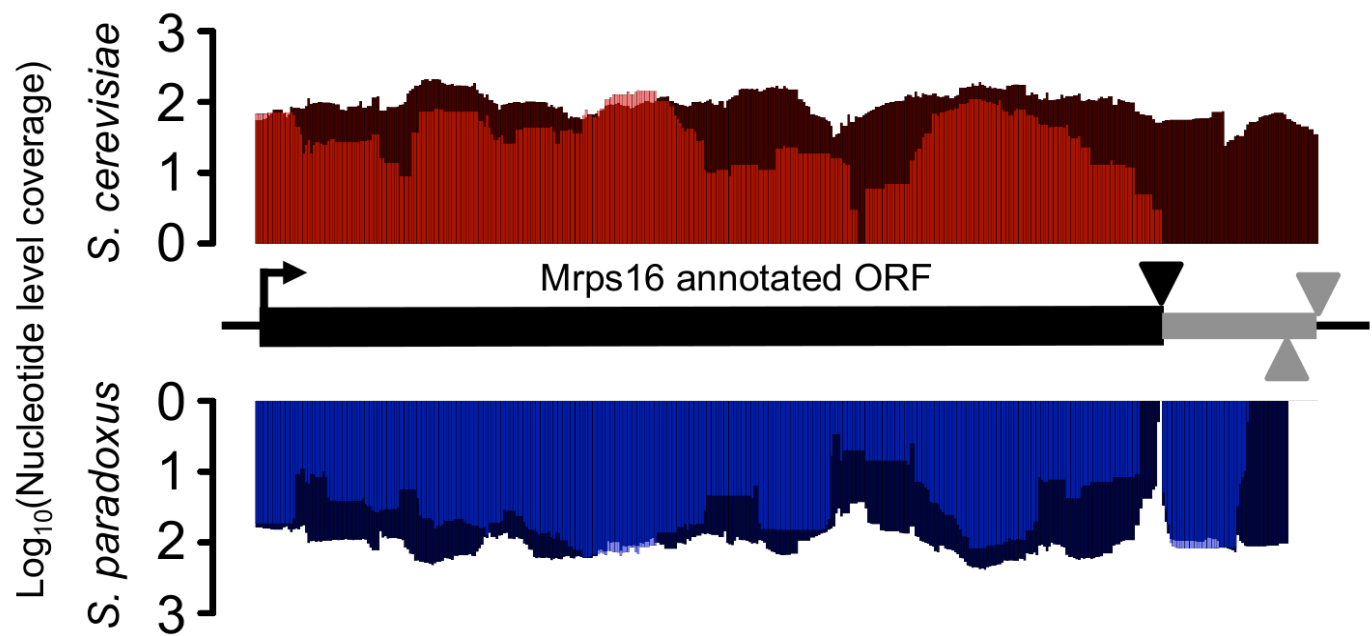

FIG 4

**SUPPLEMENTAL MATERIALS**

**Neither level of expression, nor extreme mRNA *cis* ratios explain the observed excess of opposing differences in *cis* regulation at both regulatory levels**

Two potential non-biological explanations could produce an excess of opposing instances of transcriptional versus translational *cis* divergence. First, there could be a non-linear relationship between RPKM estimates of abundance from the mRNA and Ribo fractions. For instance, if Ribo abundance was systematically underestimated for highly expressed genes, a reduced mRNA abundance in one of the parental alleles in the hybrid would lead to an increase in measured relative translational efficiency, producing a signal of opposing changes where none exists (a systematic overestimate of Ribo abundance among genes with low expression would produce a similar effect). The opposite non-linear relationships—where Ribo abundance is systematically over-estimated among highly expressed genes or under-estimated among genes with low expression—would lead to an excess of reinforcing transcriptional versus translational *cis* divergence. However, we observe that both opposing and reinforcing mRNA/translational *cis* divergence are represented across the range of mRNA expression levels (estimated from the mean RPKMs across combined *S. cerevisiae* and *S. paradoxus* hybrid replicates; Supplemental Fig. S7A). Furthermore, the distribution of mRNA expression levels for genes with opposing or reinforcing *cis* divergence are not significantly different from that of one another (Kruskal-Wallis rank sum test, p = 0.46; Supplemental Fig. S7B).

Second, we could be systematically overestimating the hybrid mRNA *cis* ratio (or underestimating the hybrid translational *cis* ratio) in some fraction of orthologs, leading to an excess of opposing divergence among orthologs with high absolute $^{Sc}/_{Sp}$ mRNA/Ribo. The opposite effect would produce an excess of reinforcing divergence. However, again the distribution of translational *cis* ratios for genes with opposing or reinforcing *cis* divergence is represented across the range absolute mRNA *cis* ratios (Supplemental Fig. S7C) and neither the distribution of absolute mRNA *cis* ratios, nor absolute translational *cis* ratios, is significantly different between the two classes (Kruskal-Wallis rank sum test, p = 0.94 and 0.75 for the mRNA and translational levels, respectively; Supplemental Fig. S7D).

**Analysis of the relationship between genomic sequence and regulatory divergence**

The ability to accurately measure ASE in hybrid or mixed parental samples is dependent on having a sufficient number of fixed sequence differences between orthologs to confidently assign short reads (or hybridize labeled samples on microarrays) to each parental allele (De Veale et al. 2012). Indeed, we observed a significant correlation between the degree of sequence divergence (measured as % divergence; see Supplemental Methods) in the CDSs of orthologs and the absolute magnitude of *cis*-regulatory divergence (Spearman's $\rho$ = 0.12 and 0.097, $p < 1 \times 10^{-15}$ and $p = 3.9 \times 10^{-9}$ for the mRNA and translational levels, respectively). In addition, we observed a slight but significantly lower % divergence among orthologs with non-significant *cis* divergence at the mRNA level (Kruskal-Wallis test, $p = 0.00070$); however, this effect disappears by removing the 380 (~10%) least divergent orthologs ($p = 0.054$), indicating that it is being driven by the most highly conserved genes. No such relationship between % divergence and our ability to detect significant *cis*-divergence at the translational level was observed ($p = 0.41$). Removal of the 380 least divergent orthologs from our dataset had no qualitative impact on our observation of an excess of opposing *cis*-divergence between regulatory levels (data not shown).

Local variability in mutation rates may lead to similar patterns of divergence in neighboring locations in the genome (e.g., promoters and their associated CDSs; Hellman et al. 2005). As expected, sequence divergence in promoter regions (defined as -200 to -1 nt relative to the transcriptional start site [TSS]), 5′ UTRs and 3′ UTRs are significantly positively correlated with divergence in their CDSs (Spearman's $\rho$ = 0.20, 0.20, and 0.21, $p < 10^{-15}$, for promoter regions, 5′ UTRs and 3′ UTRs, respectively). Therefore, in order to account for the possibility that any relationships detected in divergence at non-CDS regions spuriously reflects CDS divergence, we performed a multiple regression analysis testing for independent association between the magnitude of the absolute $^{Sc}/_{Sp}$ *cis*-ratio at either level and divergence in the promoter, the 5′ UTR, the CDS, or the 3′ UTR (see Supplemental Methods). At the mRNA level, we found that % divergence in the 5′ UTRs was significantly correlated with the absolute $^{Sc}/_{Sp}$ *cis*-ratio ($p = 0.015$), while % divergence in promoters and 3′ UTRs was not ($p = 0.41$ and $0.22$, respectively). At the translational level, only divergence in the 3′ UTR was significantly correlated when controlling for the effect of CDS (p = 0.002). However, a recent study found that translational dynamics were strongly related to nucleotide sequences in the 5′ UTR immediately adjacent to the start codon (Dvir et al. 2013). Therefore we performed the same analysis as above using only divergence of the first 50 bp of the 5′ UTR (and analyzing those UTRs that were >= 50 bp in length). In this case, the relationship between divergence in the 5′ UTR and the absolute mRNA *cis* ratio improved (p = 0.0042), and became the strongest predictor of absolute *cis* divergence at the translational level (p = 0.00069 and 0.0087, for the last 50 bp of the 5′ UTR and the 3′ UTR, respectively). These results may suggest that 5′ UTRs harbor elements that regulate either (or both) mRNA abundance and translation; however, a recent study by Pelechano et al. (2013) found that most genes of *S. cerevisiae* produce multiple isoforms with alternative TSSs. Therefore, it is also possible that our observations in the 5′ UTRs and promoters simply reflect an inability precisely define the boundaries of these elements (if precision in such boundaries exists).

**Analysis of the relationship between divergence in mRNA secondary structure and translation**

We determined the minimum free energy (MFE) in sliding 41 nucleotide windows using a 10 bp step, for the region -100 to +100 surrounding the first nucleotide of the start codon of the 3,665 orthologs analyzed in the hybrid data (see Methods in the main text) using RNAfold with default parameters (Hofacker et al. 1994). For each window, we calculated ΔMFE (MFE$_{Scer}$ – MFE$_{Spar}$) and determined its correlation with either log$_2$($^{Sc}/_{Sp}$ mRNA *cis* or translational *cis*). In this case, we found a positive correlation where reduced secondary structure (higher MFE) is associated with increased expression. We observed several windows with significant positive correlations in the translational *cis* ratio and no negative correlations, consistent with the notion that changes in secondary structure can affect translational efficiency in the expected direction (Supplemental Fig. S10). Note that in all cases, correlation coefficients are < 0.1, suggesting that ΔMFE can explain only a small fraction of the variance in translational efficiency. At the same time we observed an opposite relationship with the mRNA *cis* ratio (Supplemental Fig. S10), which could reflect a relationship between sequence composition and transcriptional dynamics in the 5′ UTR/promoter region or, alternatively the preferential sequenceability of transcripts with particular nucleotides associated with stronger secondary structure in their 5′ ends (i.e., G and C; Zheng et al. 2011). Because sequence, whether via mRNA secondary structure or not, affects sequenceability of NGS libraries (Zheng et al. 2011), it may confound results derived from computational prediction of ΔMFE. However, we note that the relationship observed in windows beginning at +30 are unique to the translational *cis* ratio, supporting an effect of secondary structure on translation at the beginning of the CDS (e.g., Tuller et al. 2011).

**No evidence that translation in 5′ UTRs is a significant determinant of *cis*-regulatory divergence in translational efficiency**

Allele-specific presence and/or translation of uORFs could provide a plausible mechanism explaining divergence of translational efficiency in *cis*. A well-studied example of this phenomenon is the *GCN4* system in *S. cerevisiae*, which represses translation of the main ORF via four uORFs under nutrient rich conditions, and activates translation in response to amino acid starvation (Hinnebush 1997). However, a recent riboprofiling analysis of yeast meiosis found that changes in translation of most uORFs were positively correlated with translation of the main ORF, indicating that the former's repressive effects are far from universal (Brar et al. 2012).

Identification of homologous uORFs between even closely related species is challenging, due both to their short lengths (the median length of annotated uORFs in the *S. cerevisiae* genome is 33 nucleotides) as well as the lack of evidence for translation or function at many potential uORFs (Ingolia et al. 2009). Therefore, we first compared patterns of upstream translation in the Ribo samples in both species using the annotated 5′ UTRs of *S. cerevisiae* as well as an equivalent length of sequence upstream of the start codon in *S. paradoxus* (via these criteria, 90% of annotated 5′ UTRs expressed in the mRNA fractions are detected in both species). The 5′ UTR with the highest average coverage between species was *GCN4*, strongly suggesting that its function in stress response remains conserved. Evidence of translation was observed in 387 and 373 5′ UTRs in *S. cerevisiae* and *S. paradoxus*, respectively (see Supplemental methods) (Supplemental Table S5). Significant translation was detected in both species in 223 5′ UTRs; very few upstream sequences (51) showed species-specific evidence of translation (i.e., reads mapping in both replicates of one species, but no reads mapping in either replicate of the other). Interestingly, orthologs with detectible 5′ UTR translation in both species are significantly over-represented for genes involved in stress response (p = 2.1 × $10^{-6}$) suggesting that the mechanism of translational repression employed by *GCN4* may not be unique. There is no significant excess of orthologs with *cis*-regulatory divergence in translational efficiency among those with detectible translation in their 5′ UTRs ($\chi^2$ = 0.09, p = 0.76). Furthermore, there is no evidence of a negative correlation between detection of significant 5′ UTR translation in one species and allele-specific translation bias favoring the other ($\chi^2$ test, p > 0.05 in all cases; note that because of the low number of reads mapping to 5′ UTRs we simply asked if the directionality of bias was the same or opposite without assigning a significance to the bias). This remains the case when restricting the analysis only to orthologs with species-specific 5′ UTR translation. While it is possible that some uORFs act in a species-specific *cis* fashion to affect translational efficiency, the small proportion of orthologs with significant upstream translation (~14%) and their lack of enrichment among orthologs with *cis* divergence in translation makes it unlikely that this is a significant mechanism explaining divergence in translational efficiency. In addition, a recent study by Pelechano and colleagues (2013) noted that many *S. cerevisiae* transcripts express alternative mRNAs that can exclude potential uORFs. Supporting their findings, we also observe that orthologs with significant 5′ UTR translation in both species and that contain uORFs that are in the 5′ UTRs of all detected transcripts (63) show significantly reduced mean hybrid translational efficiency when compared to those orthologs with upstream translation but lacking uORFs (Kruskal-Wallis rank sum test, p = 0.0059). Therefore a systematic analysis of species-specific uORF action will likely require characterizing the alternative transcriptional landscape of both species, coupled to a more thorough identification of translated uORFs using riboprofiling modified to specifically detect sites of translational initiation (e.g., Ingolia et al. 2011).

**SUPPLEMENTAL METHODS**

*Riboprofiling library construction*

The following modifications were made to the method of Ingolia (2010): Cryo-grinding of lysates was performed in a Retsch Mixer-Mill MM 301 (Retsch Technology GmbH) at maximum frequency for two 1.5 minute cycles with immersion in liquid $N_2$ before grinding, in between cycles, and following grinding. After purification of the cryo-ground lysate, RNA abundance was determined from the $A_{260}$ measured using a Nanodrop 2000c (Thermo Scientific) and 1000 µg of RNA was subjected to density gradient centrifugation for monosome isolation. Gradients were fractionated and fractions corresponding to the 80S monosome were collected using a Biocomp Instruments Gradient Station attached to a Foxy Jr Fraction Collector (Teledyne Isco). RNA was extracted from the sucrose gradient fractions using the RNeasy Mini Kit (Qiagen). Total RNA was isolated from the purified lysate using the Epicenter MasterPure™ Yeast RNA Purification Kit beginning with 500 µg of lysate RNA diluted to 125 µl using polysome lysate buffer. Following circularization, the libraries were subjected to an rRNA subtraction step as described in Brar et al. (2012).

NGS libraries were sequenced as follows: Hybrids: mRNA and Ribo fractions were each sequenced on individual lanes of a flowcell. Parents: one replicate of mRNA and both replicates of the Ribo fraction libraries were combined to approximately equal proportion and sequenced on individual lanes of a flowcell. The second replicate of the mRNA fraction for each parental strain was kept separate and sequenced on an individual lane of the flowcell in order to compare the sequence obtained from our strains to the genome assemblies and reannotate any single nucleotide polymorphisms (see below). Combined parental mRNA replicate two was subsequently generated by randomly combining 60,000,000 reads from each of the two parental replicates *in silico*.

*Iterative mapping of riboprofiling reads*

Reads were mapped according to the method of Ingolia et al. (2010) as follows: Beginning with the individual parental mRNA samples, we first excluded any reads that, when trimmed to 23 bases from the 5′ end, mapped to the complete rDNA sequence of *S.*

*cerevisiae* allowing 3 mismatches and a maximum of 20 mapping locations using Bowtie version 0.12 (Langmead et al. 2009). Remaining reads from the parental mRNA samples were mapped to their respective genomes allowing no multimappers, and a single mismatch. Mapping reads were filtered such that no more than 30 bp (31 bp if the 3′ most base ended with an A), and no less than 27 bp (28 if the 3′ most base was an A) from the 5′ end of the read mapped uniquely. These were used to reannotate the genome assemblies by identifying nucleotides that were overlapped by at least 10 reads differing at a specific nucleotide with the reference genomes at a frequency >= to 0.8, and that did not introduce nonsense mutations in annotated genes (the absence of true nonsense mutations was confirmed in the five cases where substitutions were detected by the presence of abundant Ribo fraction read coverage 3′ of the putative stop codon). This identified 239 and 605 differences between our data and the *S. cerevisiae* and *S. paradoxus* assemblies, respectively. Replicate reads from all samples were then mapped to a concatenation of the updated assemblies, as above, but allowing no mismatches.

*S. cerevisiae* and *S. paradoxus* show sufficient divergence at the nucleotide level (~5%) that the 27-30 nt RNA fragments produced by the riboprofiling protocol mapped uniquely to most genomic regions (~87% when accounting for non-unique regions both within and between the two genomes). In order to analyze only unique mapping nucleotides, non-unique mapping nucleotides were identified by truncating each of the species' genomes into overlapping 27 bp fragments in single-base increments along each chromosome. These fragments were then mapped back to the concatenation of the two species' genomes using Bowtie allowing no mismatches and removing all locations spanned by reads mapping to more than a single location (multi-mapping reads).

*Identification of high-confidence Scannell et al. (2011) orthologs*

From the list of genes orthologous between *S. cerevisiae* and *S. paradoxus*, we identified those that in both species a) began with an ATG and terminated in a canonical stop codon (TAA, TAG, TGA), b) had a sequence length that was divisible by three, c) lacked in-frame stop codons, d) lacked any 'N' nucleotides in their genomic sequence, e) were annotated as either possessing or lacking introns in both species, and f) possessed at least 100 uniquely mappable nucleotides. Furthermore, we required that the lengths of both orthologs be within 50% of one another and excluded genes that were annotated as having a different number of introns in the Scannell et al. (2011) *S. cerevisiae* annotation than in the Saccharomyces Genome Database (SGD; Cherry et al. 2012) annotation available as of 14 August 2012. The Scannell et al. (2011) annotation provides spliced, processed mRNA sequences for each annotated gene, however the positions of introns are not indicated in the genomic annotation files. These were obtained by using BLAT (Kent 2002) to map each species' mRNA to its respective genome. Intron flanking segments of 84 of the 105 intron containing mRNAs mapped uniquely and were retained for analysis. Finally we eliminated orthologs for LYS2 (Scer_2.299) as its knockout in the *S. cerevisiae* parental strain was used as a selectable marker as well as CTR3 (Scer_12.598) as it is interrupted by a Ty2 transposon in BY strains of *S. cerevisiae*. Our final analysis set contained 4,640 orthologs between the two parental species (Supplementary Table S1).

*Applying the method of Bullard et al. (2010) to detect significant mRNA ASE*

The test involves resampling, with replacement, the base-level read coverage of each parental allele 10,000 times, under two conditions: 1) using the *S. cerevisiae* marginal nucleotide frequencies ($\pi_c = \pi_c[A], \pi_c[C], \pi_c[G], \pi_c[T]$) and the *S. cerevisiae* length, $L_c$, and 2) using the *S. paradoxus* marginal nucleotide frequencies $\pi_p$ and the *S. paradoxus* length, $L_p$. A started $\log_2$ ratio (total base level coverage from $\pi_c, L_c + 1$ / total base level coverage from $\pi_p, L_p + 1$), denoted as $\log_2(^{Sc+1}/_{Sp+1})$, was obtained from each resampling representing the variation in $\log_2(^{Sc+1}/_{Sp+1}$ mRNA) ratios expected between alleles due only to differential base frequencies and length. The two null distributions (one per allele) were compared against the observed started $\log_2(^{Sc+1}/_{Sp+1}$ mRNA) ratio in order to obtain a two-tailed p value based on how often the observed ratio was outside of the bounds of the null distribution. If both replicates agreed in the direction of parental bias, we retained the least significant p-value in either replicate as a measure of the significance of differential expression.

*Detecting significant trans-regulatory divergence using parental data*

For the purpose of analyzing *trans*-divergence, we focused on those orthologs with a minimum of 100 reads mapping among both alleles within all replicate mRNA fraction. Furthermore, we removed any ortholog identified as being differentially expressed among different mating types (18 orthologs; Galitski et al. 1999), and/or ploidy levels (35 orthologs; Wu et al. 2010) as the *S. cerevisiae* strain BY4716 is haploid while the *S. paradoxus* CBS432 strain is diploid. Ploidy level has previously been shown to have no significant effect on estimates of *trans*-regulatory divergence between these species (Tirosh et al. 2009). The 3,634 remaining orthologs were used to test for significant *trans* regulatory divergence at both levels using the same approach as outlined above, modified as follows: For the test of significant *trans* regulatory divergence in mRNA abundance, we sought to reject the null hypothesis that $\log_2(\text{parental }^{Sc+1}/_{Sp+1}$ mRNA) was not significantly different from $\log_2(\text{hybrid }^{Sc+1}/_{Sp+1}$ mRNA). Therefore, we resampled the CDS base-level coverage of the *S. cerevisiae* allele using $\pi_c$ and $L_c$ and the *S. paradoxus* allele using $\pi_p$ and $L_p$ 10,000 times in the each replicate of the parental mRNA fraction. Each resampling was used to generate a distribution of $\log_2(\text{parental }^{Sc+1}/_{Sp+1}$ mRNA) ratios, which takes into account the variability in read coverage across each allele. These distributions were then compared to the mean observed $\log_2(\text{hybrid }^{Sc+1}/_{Sp+1}$ mRNA) to generate a p-value. The same resampling was then repeated reciprocally in each hybrid mRNA fraction replicate, which was then compared to mean observed $\log_2(\text{parental }^{Sc+1}/_{Sp+1}$ mRNA). As above, if the directionality of difference agreed among all individual replicate comparisons (i.e., both observed replicate $\log_2(\text{hybrid }^{Sc+1}/_{Sp+1}$ mRNA) had to agree in direction when compared to both observed replicate $\log_2(\text{parental }^{Sc+1}/_{Sp+1}$ mRNA), the least significant of the four p-values was retained.

For the test of significant *trans* regulatory divergence in translation, we sought to reject the null hypothesis that $\log_2(\text{parental }^{Sc+1}/_{Sp+1}$ Ribo) was not significantly different from sum of $\log_2(\text{hybrid }^{Sc+1}/_{Sp+1}$ Ribo) and $\log_2(\text{parental }^{Sc+1}/_{Sp+1}$ mRNA). Therefore, we resampled the CDS base-level coverage of the *S. cerevisiae* allele using $\pi_c$ and $L_c$ and the *S. paradoxus* allele using $\pi_p$ and $L_p$ 10,000 times in each replicate of the parental Ribo fraction and compared the resulting distributions the sum of the mean observed

$\log_2(\text{hybrid }^{Sc+1}/_{Sp+1}\text{ Ribo})$ and $\log_2(\text{parental }^{Sc+1}/_{Sp+1}\text{ mRNA})$ to generate a p-value. The same resampling was then repeated reciprocally to generate two permuted distributions where each replicate permutation of $\log_2(\text{hybrid }^{Sc+1}/_{Sp+1}\text{ Ribo})$ was summed with one or the other replicate permutation of $\log_2(\text{parental }^{Sc+1}/_{Sp+1}\text{ mRNA})$ with equal probability. If the directionality of difference between the both $\log_2(\text{parental }^{Sc+1}/_{Sp+1}\text{ Ribo})$ and the mean summed ratios agreed, the least significant of the four p-values was retained. Differences significant at 5% FDR were retained.

*Criteria for identification of candidate C-terminal extensions*

We combined the two replicate Ribo fractions in the hybrids and parents for the purpose of assessing if a candidate C-terminal extension was translated (however, Supplemental Table S4 indicates if reads were detected in both replicates). A number of different criteria were used to assess the potential validity of 3′ readthrough: Readthrough was considered species-specific if ≥ 5 reads mapped to the extension in both the combined hybrid and combined parental replicates in one species but < 3 mapped in either of the combined samples in the other species. In order to consider read through conserved between species, in addition to meeting the above criteria, we required 1) the presence of ≥ 5 mapping reads in one species and ≥ 3 reads in the other species in both the combined hybrid and the combined parental replicates 2) the absence of frame shifting indels in the aligned C-terminal extensions. Conserved read through candidates were then scored as 'Good' if the ratio of non-synonymous substitutions per non-synonymous site (Ka) to synonymous substitutions per synonymous site (Ks) was < 0.8 as determined by aligning the putative extension using DIALIGN-TX version 1.0.0 (Subramanian et al. 2005) and RevTrans version 1.4 (Wernersson and Pedersen 2003), followed by KaKs Calculator version 2.0 using the 'NG' method (Zhang et al. 2006), or 'Poor' if it was ≥ 0.8 and/or translation was detected using the conserved criteria in only the hybrid or parental combined replicates. In cases where the Ks was 0, candidates were considered 'Good' if they experienced ≤ 2 non-synonymous substitutions.

*Analysis of translation in 5′ UTRs*

Ribo samples were mapped to the regions identified as *S. cerevisiae* 5′ UTRs (see Methods in the main text) by Nagalakshmi et al. (2008) and the sequence of an equivalent length upstream of the annotated AUG codon of *S. paradoxus* orthologs. We required at least five reads mapping to at least one species in both hybrid replicates to classify a 5′ UTR as translated in a single species, and at least five reads mapping in both replicates of both species to be translated in both. For the analysis of Pelechano et al. (2013) uORFs, we obtained their list of genes in *S. cerevisiae* whose uORFs were upstream of the main ORF in all transcripts detected.

**SUPPLEMENTAL REFERENCES**

# SUPPLEMENTAL TABLES

**Supplemental Table S2**. Overview of the location within annotated transcripts where reads from each fraction and replicate map. Reads were assigned to each feature based on the mapping location of their 5′ ends (see Methods). The number of reads mapping to each feature is indicated along with their proportion (Prop.) calculated as the fraction of reads mapping to the feature divided by the fraction of total mappable bases in that feature. Mapping locations of the mRNA fractions are 3′ biased as expected from RNA-Seq data. The Ribo fractions are more strongly biased towards reads mapping in the CDS, again as expected from previous ribosome profiling studies (e.g., Ingolia et al. 2009). We note the inconsistent number of reads mapping to 3′ UTRs in the hybrid Ribo fraction may affect our ability to identify C-terminal peptide extensions (see main manuscript), therefore the all candidates for C-terminal extensions with reads derived from both fractions are indicated in Supplementary Table 4.

| Sample | Rep | CDS 10,403,087 Reads | Prop. | Introns 16,741 Reads | Prop. | 5′ UTR 511,468 Reads | Prop. | 3′ UTR 970,922 Reads | Prop. |
|---|---|---|---|---|---|---|---|---|---|
| Hybrid mRNA Fraction | 1 | 13,830,192 | 0.99 | 9,309 | 0.42 | 370,715 | 0.54 | 1,705,144 | 1.31 |
| | 2 | 13,970,968 | 0.98 | 10,278 | 0.45 | 416,530 | 0.60 | 1,853,799 | 1.40 |
| Hybrid Ribo Fraction | 1 | 7,923,111 | 1.14 | 148 | 0.01 | 32,562 | 0.09 | 28,057 | 0.04 |
| | 2 | 5,341,475 | 1.09 | 286 | 0.04 | 35,249 | 0.15 | 239,745 | 0.52 |
| Parental mRNA Fraction | 1 | 9,528,924 | 0.99 | 8,279 | 0.54 | 269,978 | 0.57 | 1,150,185 | 1.29 |
| | 2 | 13,723,361 | 0.98 | 12,402 | 0.55 | 341,516 | 0.50 | 1,865,146 | 1.43 |
| Parental Ribo Fraction | 1 | 7,454,237 | 1.12 | 193 | 0.02 | 39,329 | 0.12 | 121,750 | 0.20 |
| | 2 | 15,488,748 | 1.13 | 252 | 0.01 | 83,251 | 0.12 | 137,296 | 0.11 |

**SUPPLEMENTAL FIGURES**

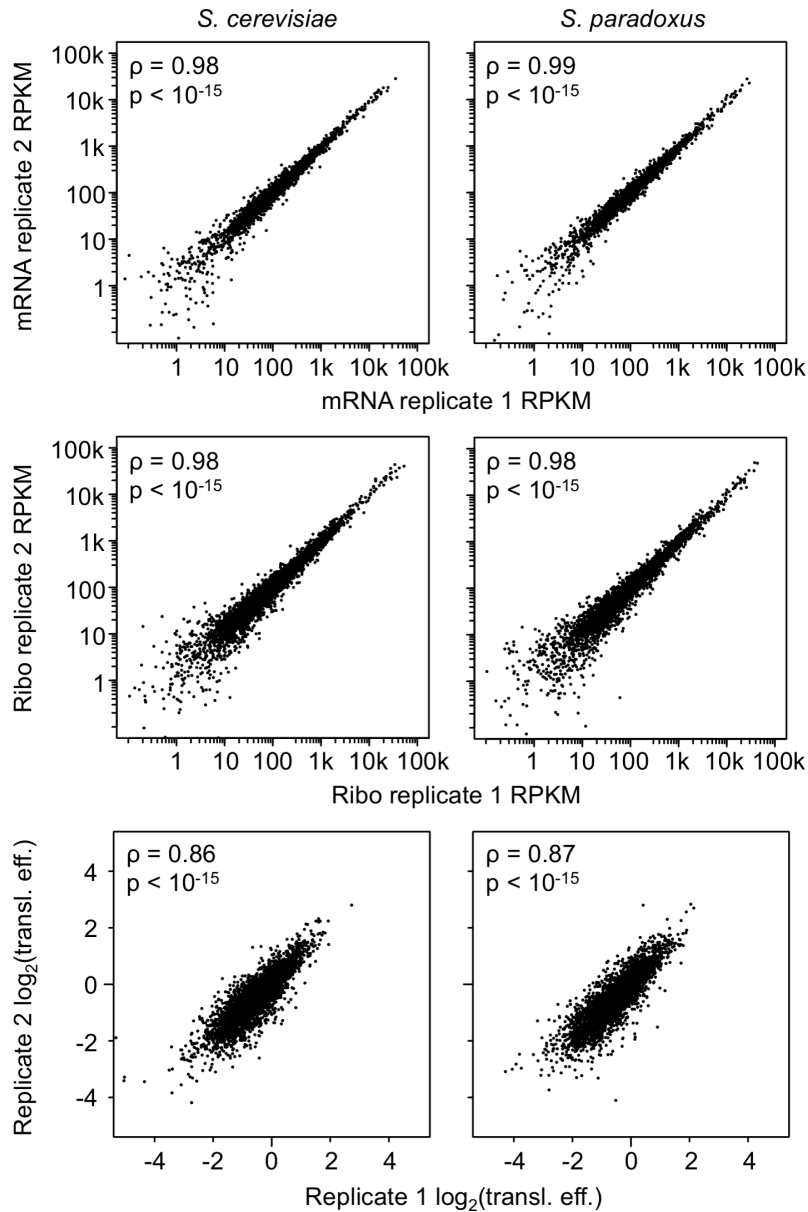

**Supplemental Figure S1.** Comparisons of hybrid biological replicate allele-specific RPKM abundance estimates for all 4,640 orthologs. *S. cerevisiae* RPKMs are shown on the left while *S. paradoxus* estimates are on the right. The mRNA fraction (A) is shown above the Ribo fraction (B). C) Comparison between hybrid biological replicates of the estimated translational efficiencies for the 3,665 orthologs with sufficient coverage to test for significant *cis*-regulatory divergence at both regulatory levels. Spearman's correlation coefficients ($\rho$) indicate that all abundance measurements are highly reproducible. Transl. eff: Translational efficiency.

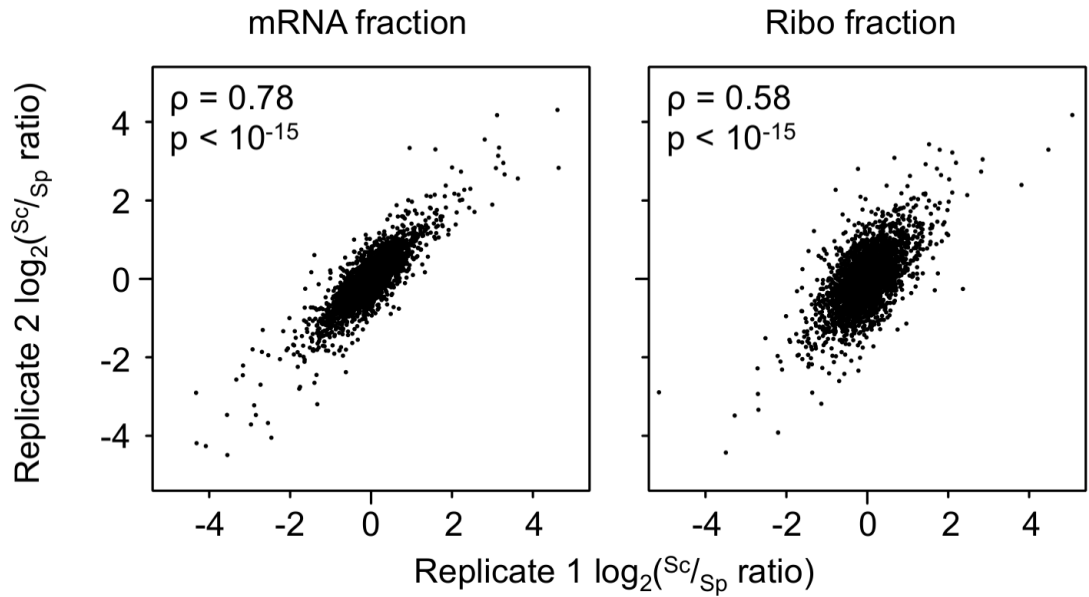

**Supplemental Figure S2.** Comparison of the estimated $^{Sc}/_{Sp}$ allelic ratios for the hybrid mRNA and Ribo fractions. Spearman's correlation coefficients (ρ) are shown in each panel. mRNA fraction $^{Sc}/_{Sp}$ ratios estimates are more reproducible, likely owing to both the greater number of mapping reads obtained from these fractions, and the more even distribution of coverage along the CDS of transcripts (Ingolia et al. 2009).

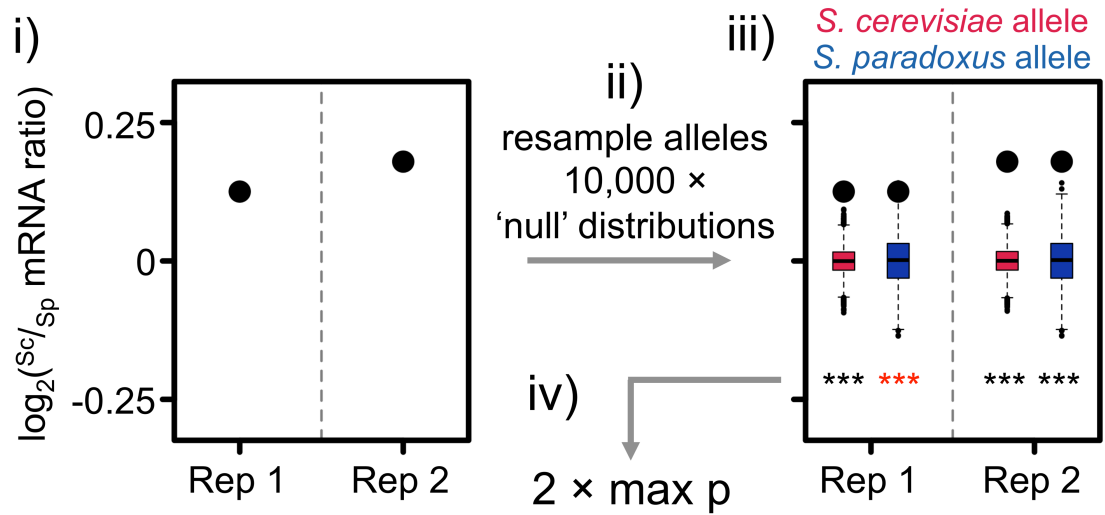

**Supplemental Figure S3.** Detection of significant divergence in mRNA abundance using the resampling approach of Bullard et al. (2010). The test is based upon rejecting the null hypothesis that the mRNA $^{Sc}/_{Sp}$ ratios are not significantly different from one. (i) The observed mRNA $^{Sc}/_{Sp}$ ratios (black circles) were obtained directly from the replicate mRNA fractions. (ii) In each fraction, the base-level coverage of each allele is resampled with replacement first using the *S. cerevisiae* marginal nucleotide frequencies and length, then using the *S. paradoxus* marginal nucleotide frequencies and length. As the same allele is resampled using the base composition and length parameters from both alleles, the expected log$_2$ ratio should be near 0 with any deviation capturing the expected inter-allelic variation due only to base composition, length differences, and read coverage. This resampling was performed 10,000 times, (iii) generating a distribution of 'null' ratios for each allele in each fraction (*S. cerevisiae*, red boxplots; *S. paradoxus*, blue boxplots). The ratio within each replicate was compared to the null distributions generated from each allele within the same replicate, for which a two-tailed p-value was calculated (note that the circles indicating the ratios in panel iii have been drawn over the permuted distributions of each allele for ease of comparison to the 'null' distributions). If all comparisons agreed in the parental direction of allelic bias (in the above example, *S. cerevisiae*), then (iv) the highest p-value (least significant as indicated by the red asterisks) was used as the representative for the test.

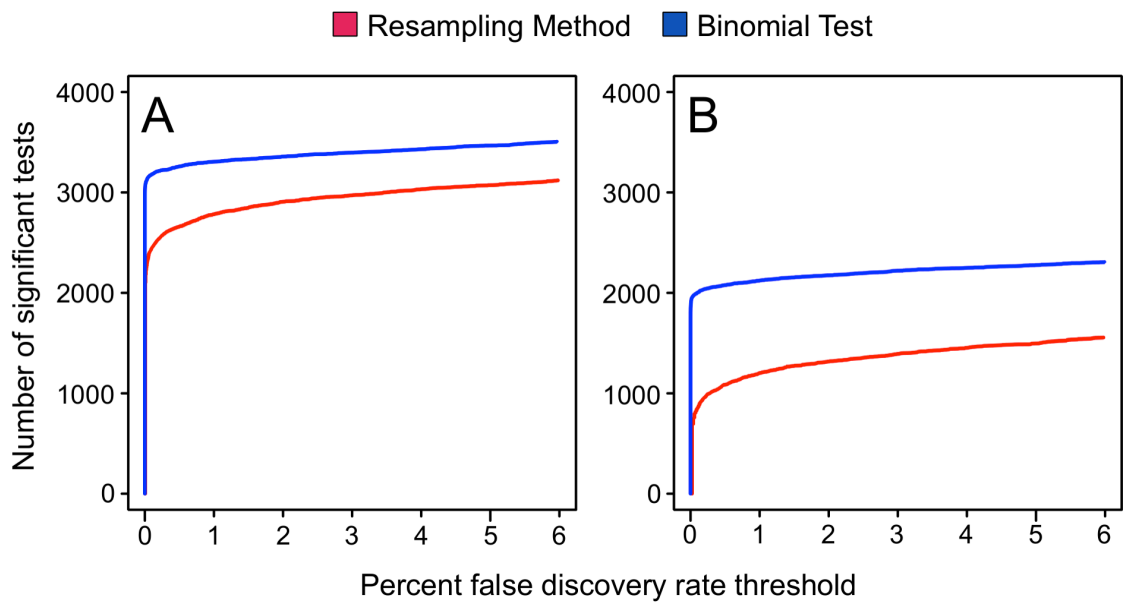

**Supplemental Figure S4.** Comparison between the number of genes showing significant regulatory divergence at increasing FDR thresholds between the resampling method as implemented in this study and the binomial test performed on the same data. The resampling based approach used in the current study is more conservative than the binomial test at both the mRNA (A) and translational (B) levels. However, this was more pronounced in the latter, as the resampling approach takes into account the increased variance in read coverage distribution in the Ribo fraction. Curves were generated using the 'qvalue' package in R (Storey 2002).

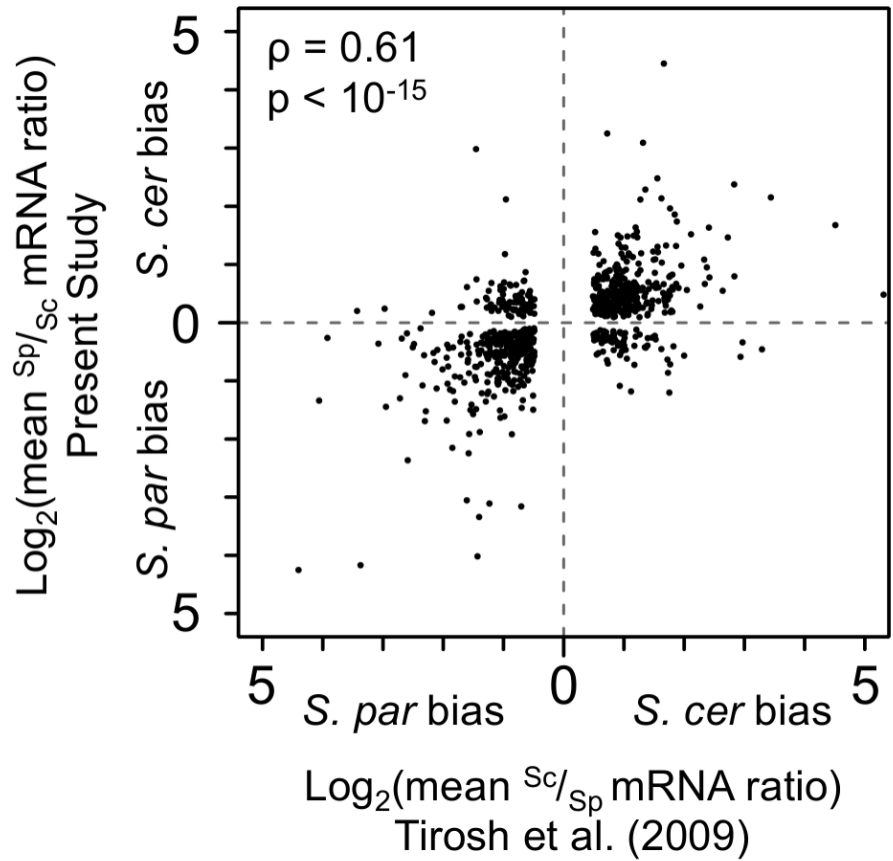

**Supplemental Figure S5.** Comparison of the results of the present study to the transcriptional *cis*-regulatory divergence estimated from Tirosh et al. (2009). Estimates of the degree of bias among genes showing significant *cis*-regulatory divergence in the transcriptional fraction agree well with the microarray-based analysis of transcriptional regulatory divergence in these species despite differences in the techniques employed (Spearman correlation coefficient in estimated $^{Sc}/_{Sp}$ mRNA ratio, $\rho = 0.61$, $p < 10^{-15}$).

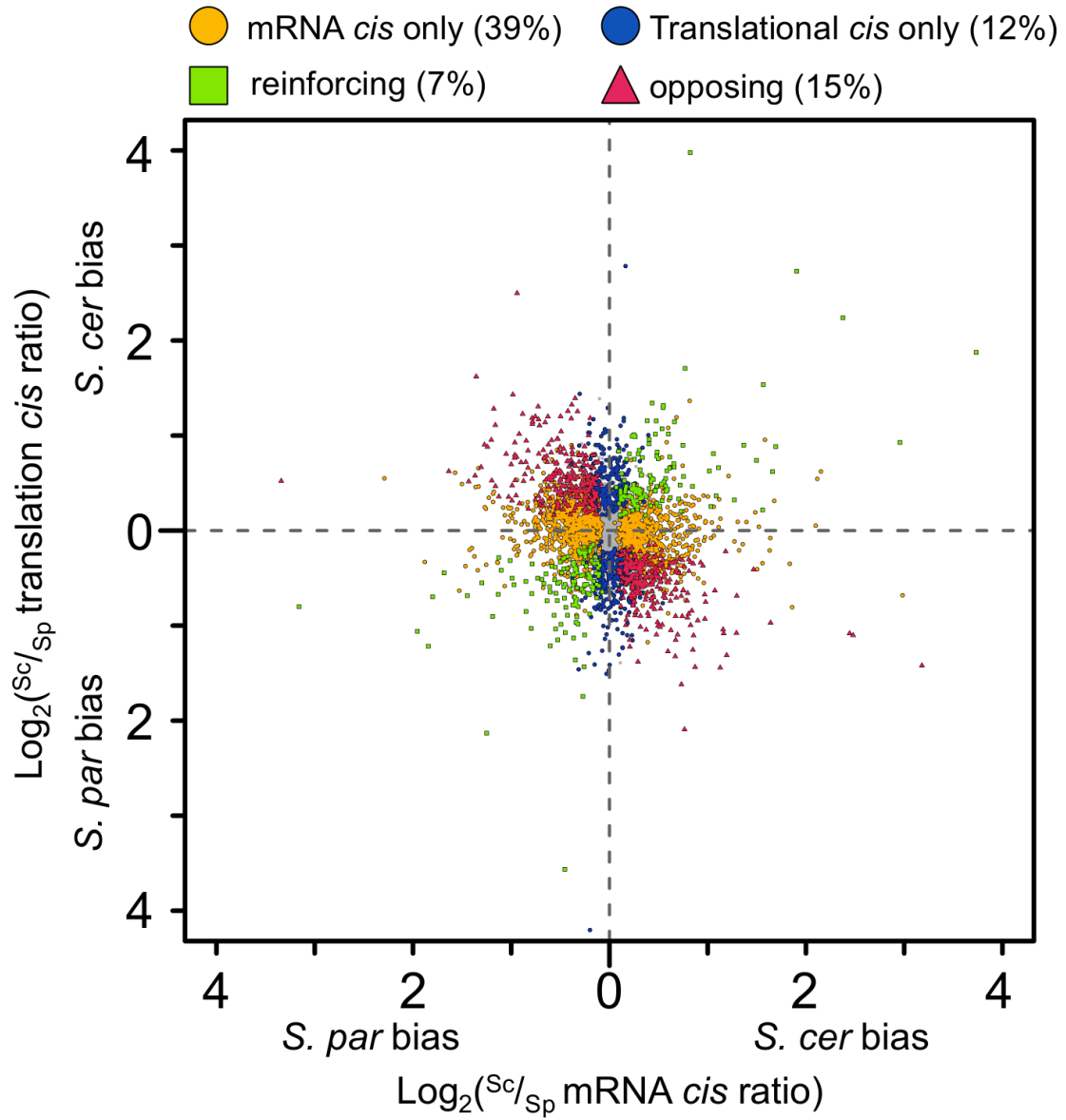

**Supplemental Figure S6.** Reproduction of Fig. 2A with the range of axes expanded to show the position of all 3,665 orthologs. *S. cer*, *S. cerevisiae*; *S. par*, *S. paradoxus*.

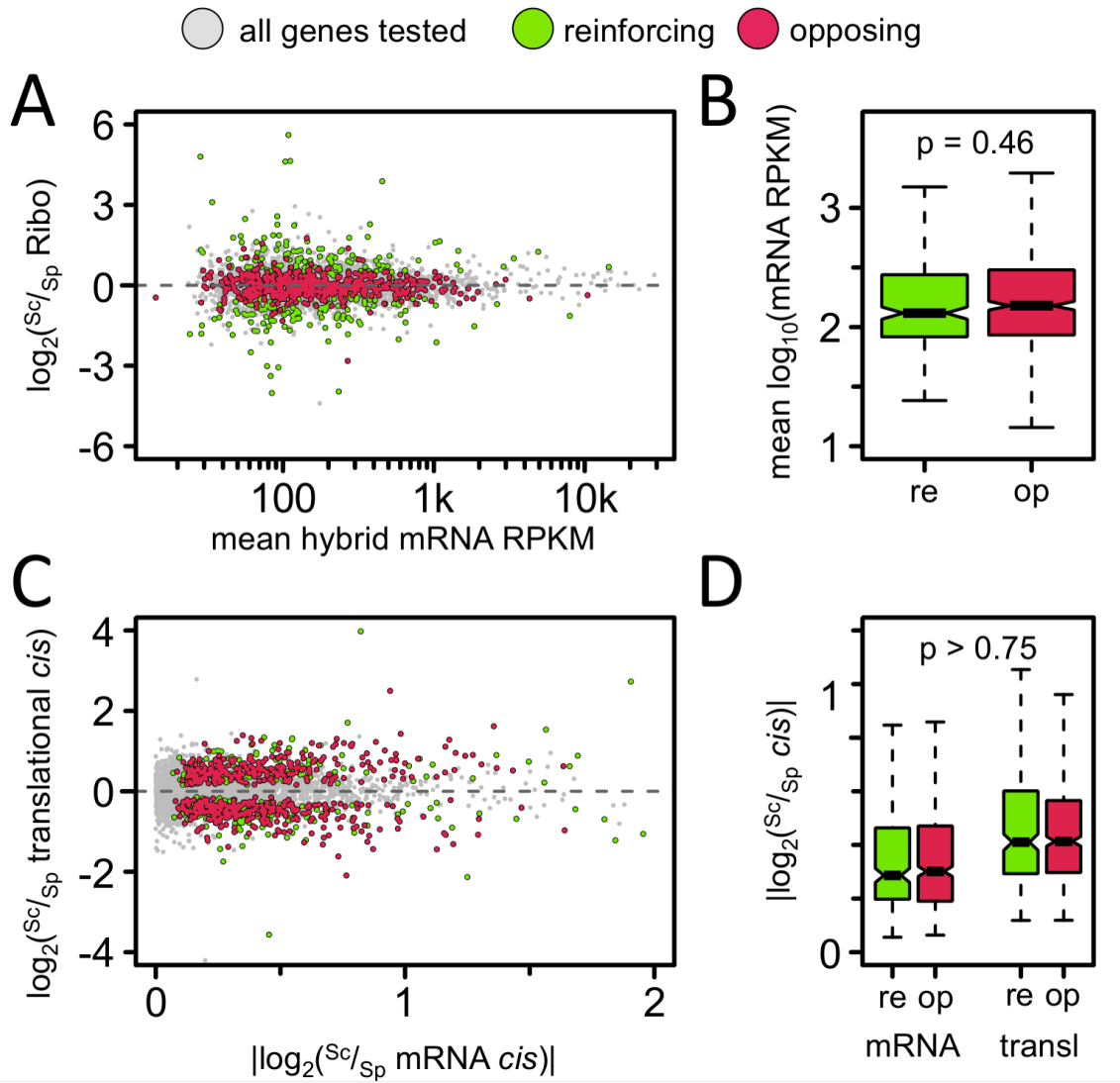

**Supplemental Figure S7.** No systematic biases are observed in mRNA expression levels or magnitudes of the *cis* ratios of orthologs with opposing or reinforcing mRNA vs. translational *cis* divergence. (A) Scatterplot of mean transcriptional RPKM (across both species and all replicates) vs. $\log_2(^{Sc}/_{Sp}$ Ribo mRNA) of orthologs with opposing (red) or reinforcing (green) mRNA vs. translational *cis* divergence. All genes tested are shown in grey. As can be seen, both types of divergence are observed across the range of expression levels. (B) Boxplot of the distribution of mRNA RPKMs for genes with reinforcing (re) or opposing divergence (op). The distributions are not significantly different (Kruskal-Wallis rank sum test, p = 0.46). (C) Scatterplot of absolute $^{Sc}/_{Sp}$ mRNA *cis* vs. $\log_2(^{Sc}/_{Sp}$ translational *cis*) of the same categories. Again, as can be seen, both types of divergence are observed across the range of *cis* magnitudes and (D) neither the distributions $|\log_2(^{Sc}/_{Sp}$ mRNA *cis*)| nor $|\log_2(^{Sc}/_{Sp}$ translational *cis*)| are significantly different among opposing vs. reinforcing orthologs (p = 0.94 and 0.75, for the mRNA and translational [transl] levels, respectively).

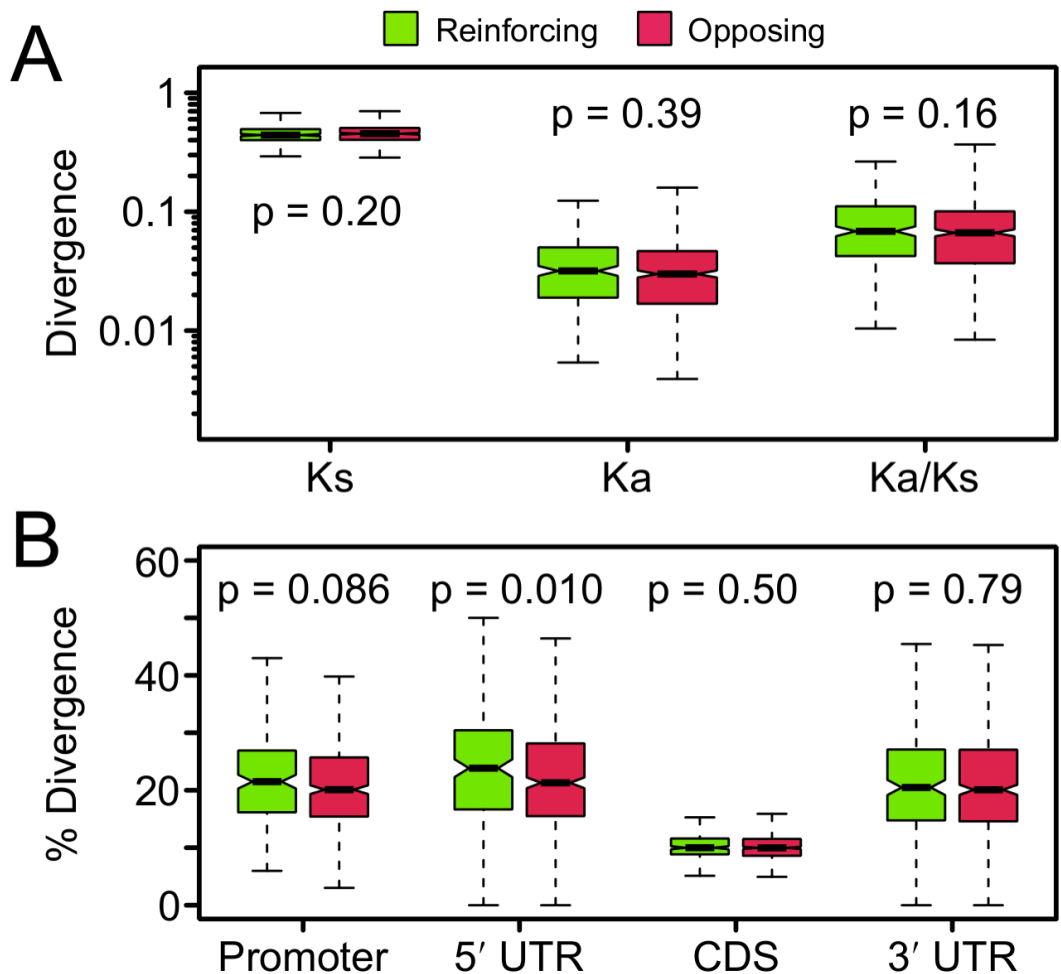

**Supplemental Figure S8.** Boxplot comparing (A) substitution rates in the CDSs or (B) levels of nucleotide divergence of orthologs with reinforcing (green) or opposing (red) *cis*-regulatory divergence between regulatory levels. Promoters are defined as 200 bp upstream of the TSS. P-values for the comparisons between categories using Kruskal-Wallis rank sum tests are shown above or below each category. The only significant comparison is a slightly reduced level of nucleotide divergence in the 5′ UTRs of orthologs with opposing divergence.

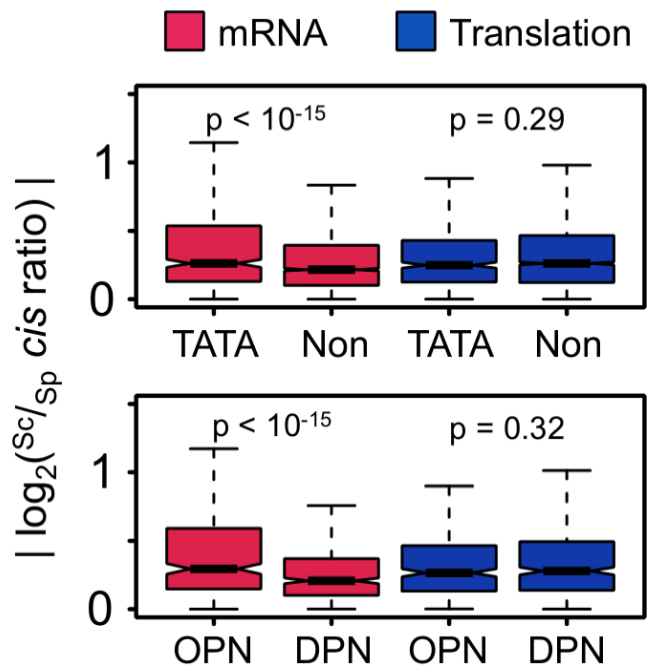

**Supplemental Figure S9.** Reproduction of Fig. 3C without controlling for the effect of the presence of OPNs in the TATA comparison, and vice versa. TATA and OPN containing genes are significantly more divergent in absolute *cis* ratio only at the mRNA level. Kruskal-Wallis rank sum test p-values are shown above each class. TATA, TATA-box containing promoter; Non, TATA-less promoter; OPN, occupied proximal-nucleosome; DPN, depleted proximal-nucleosome.

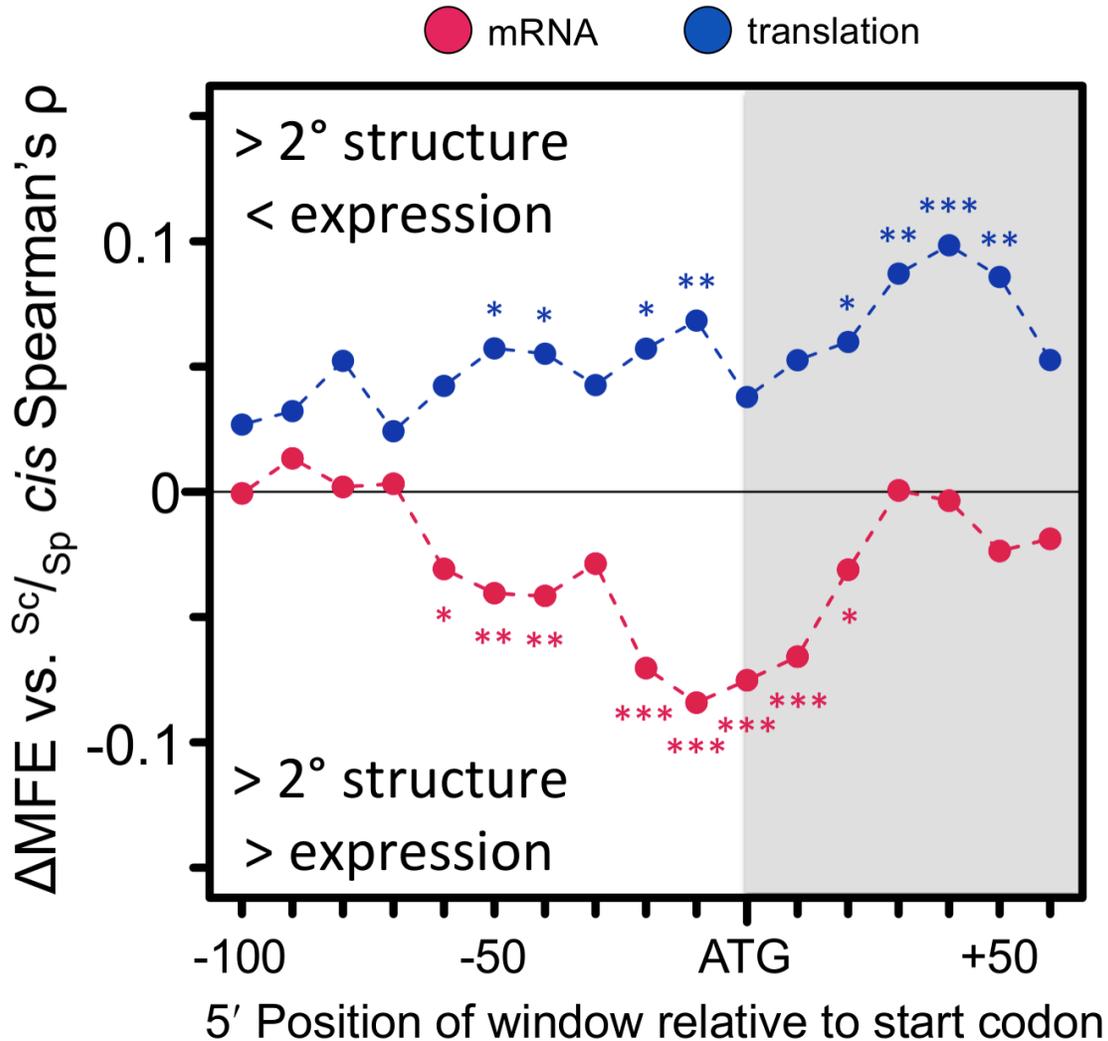

**Supplemental Figure S10.** Spearman's correlation coefficient (ρ) between change in computed minimum free energy (ΔMFE) between orthologs and *cis* ratio for 41 nucleotide windows spanning the region -100 to +100 nucleotides upstream of the start codon for the 3,665 orthologs tested for significant divergence at both levels. ρ is polarized such that lower MFE (more secondary structure) is associated with lower expression. The relationship with $^{Sc}/_{Sp}$ mRNA *cis* is shown in red, and $^{Sc}/_{Sp}$ translational *cis* in blue. Windows showing significant correlations are indicated by asterisks where * indicates $p < 0.05$, ** indicates $p < 0.01$, and *** indicates $p < 0.001$. Windows beginning in the CDS are shaded to aid in visualization. All translational correlations are in the expected direction if secondary structure hampers ribosomal access to the start codon; however, the opposite relationship is seen at the mRNA level. The relationship observed in windows beginning at +30 are unique to the translational *cis* ratio, supporting an effect of secondary structure on translation in this region (e.g., Tuller et al. 2011).

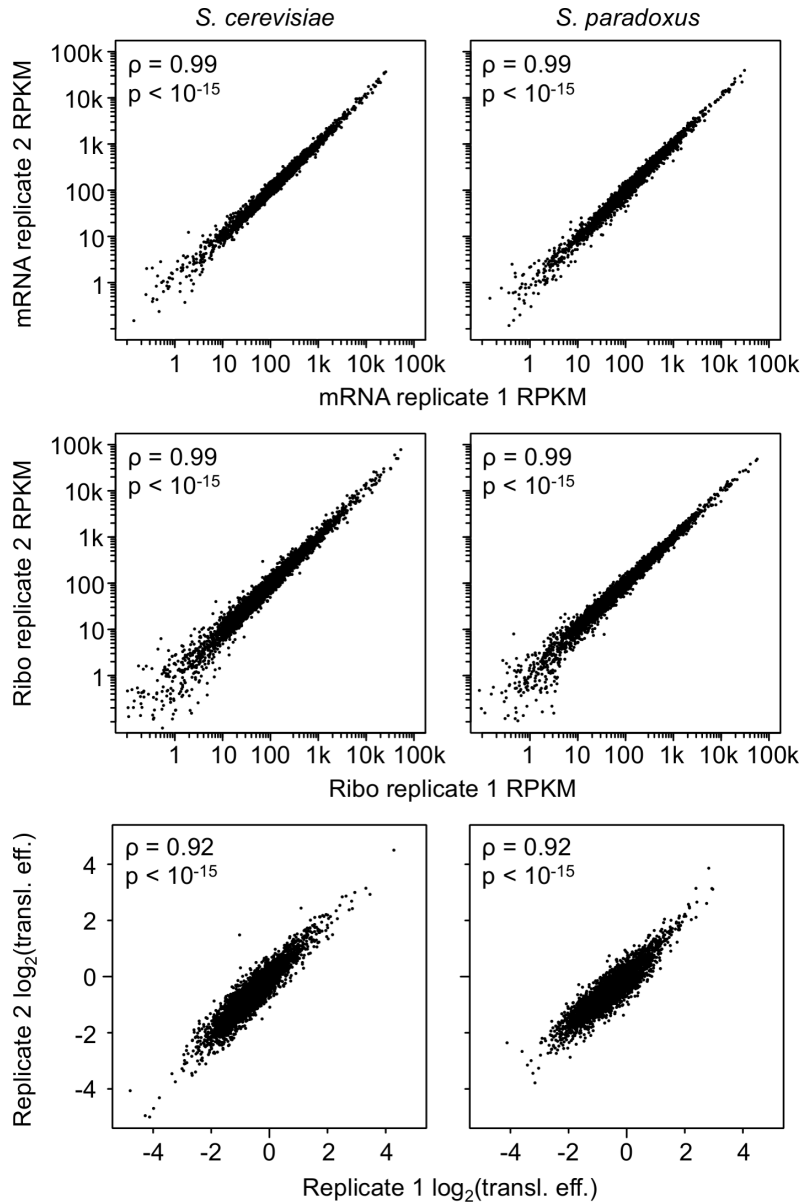

**Supplemental Figure S11.** Comparisons of parental biological replicate ortholog RPKM abundance estimates for all 4,640 orthologs. *S. cerevisiae* RPKMs are shown on the left while *S. paradoxus* estimates are on the right. The mRNA fraction (A) is shown above the Ribo fraction (B). C) Comparison between parental biological replicates of estimated translational efficiencies for the 3,634 orthologs with sufficient coverage to test for significant *cis* and *trans* regulatory divergence at both regulatory levels. Spearman's correlation coefficients (ρ) indicate that all abundance measurements are highly reproducible. The higher correlations observed for the parental data may reflect the generally greater number of reads obtained from these libraries (Supplemental Table S2). Transl. eff: Translational efficiency.

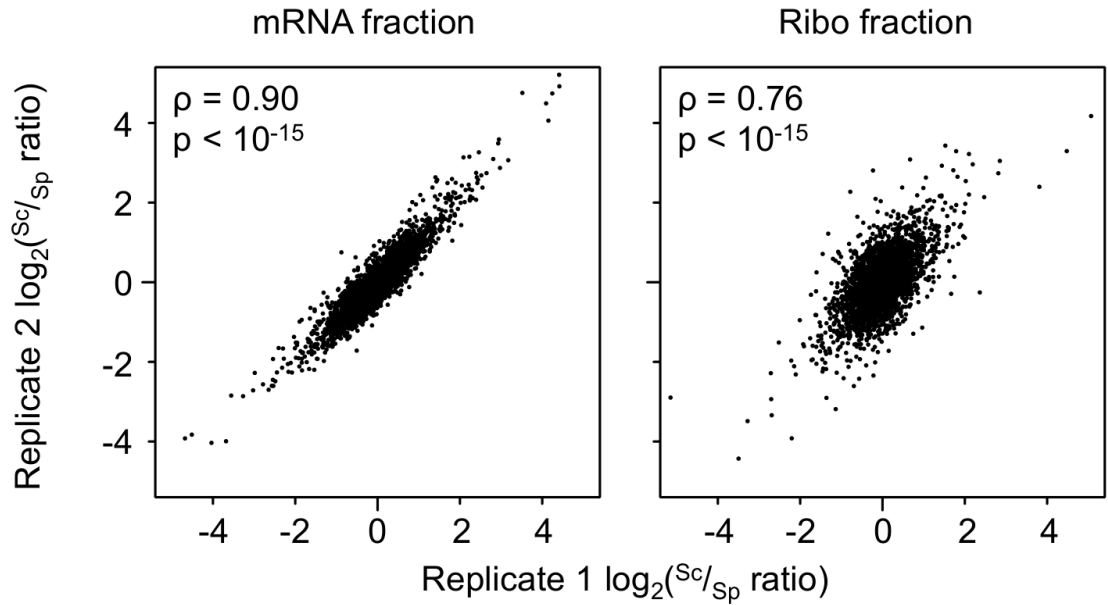

**Supplemental Figure S12.** Comparison of the estimated $^{Sc}/_{Sp}$ ortholog ratios for the parental mRNA and Ribo fractions. Spearman's correlation coefficients (ρ) are shown in each panel. As was the case with the hybrid data, mRNA fraction $^{Sc}/_{Sp}$ ratios estimates are more reproducible, likely owing to both the greater number of mapping reads obtained from these fractions, and the more even distribution of coverage along the CDS of transcripts (Ingolia et al. 2009).

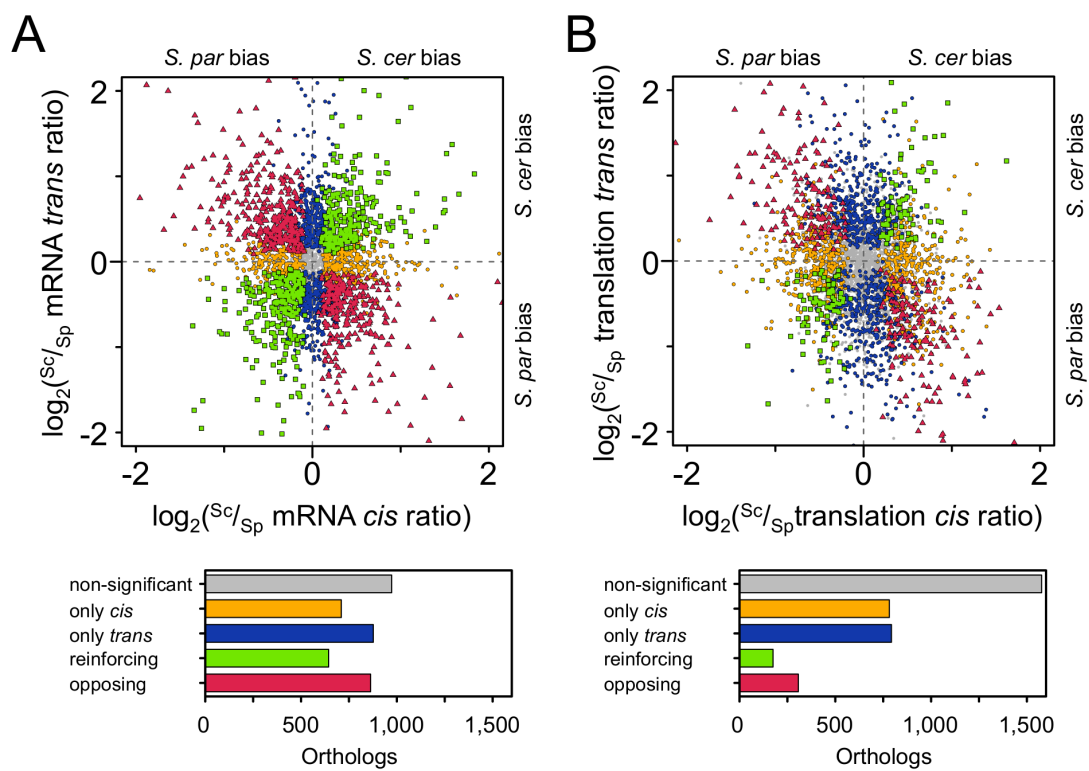

**Supplemental Figure S13.** (A) The relationship between *cis*-regulatory and *trans*-divergence at the mRNA level (all plotted $^{Sc}/_{Sp}$ ratios are the mean of the two biological replicates). While significantly more orthologs show divergence only in *trans* (blue circles) as compared to *cis* (orange circles; $\chi^2 = 14.3$, p = 0.00016), overall there is no significant excess of either type of divergence ($\chi^2 = 3.7$, p = 0.054). As was the case for *cis* divergence between regulatory levels, we observed an excess of opposing (red triangles) as compared to reinforcing (green boxes) divergence among the two regulatory mechanisms. (B) As above, but for the translational level. No significant differences are observed in the number of orthologs with significant *cis* vs. *trans* divergence, nor in those with reinforcing vs. opposing divergence ($\chi^2 = 0.030$ and 0.049, p = 0.083 and 0.86, respectively; see Supplemental Fig. S15).

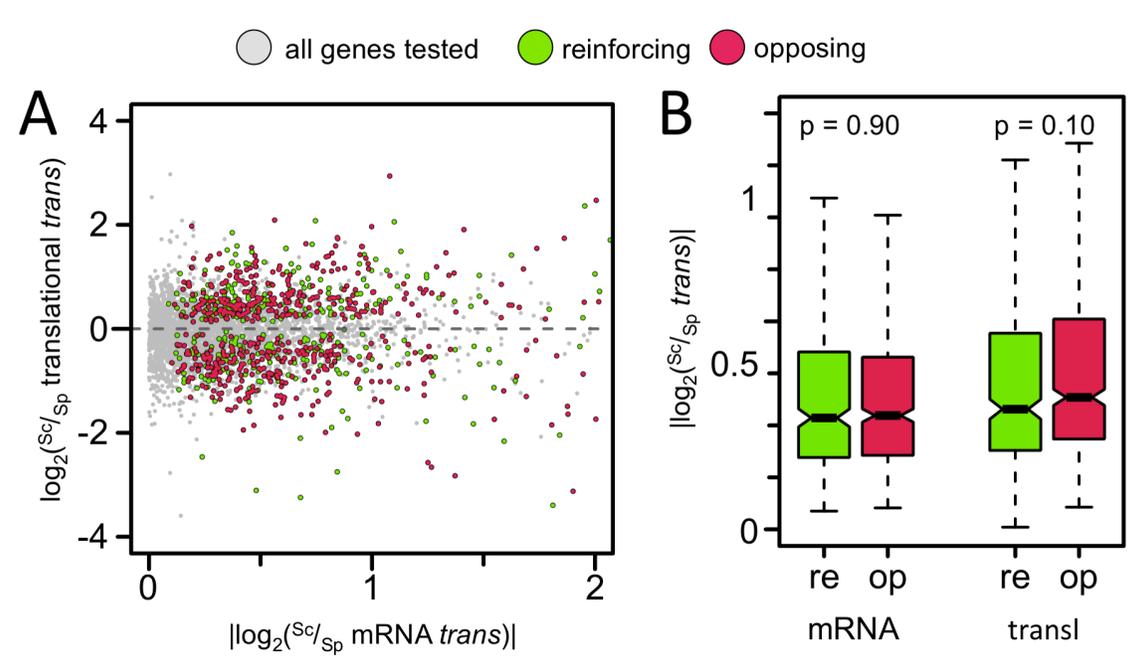

**Supplemental Fig S14.** (A) Scatterplot of $|\log_2(^{Sc}/_{Sp}\ \text{mRNA}\ trans)|$ vs. $\log_2(^{Sc}/_{Sp}$ translational $trans$) of orthologs with reinforcing or opposing $trans$ divergence across regualtory levels. (B) Neither the distributions $|\log_2(^{Sc}/_{Sp}\ \text{mRNA}\ trans)|$ nor $|\log_2(^{Sc}/_{Sp}$ translational $trans)|$ are significantly different among opposing vs. reinforcing orthologs ($p = 0.90$ and $0.10$, for the mRNA and translational [transl] levels, respectively).

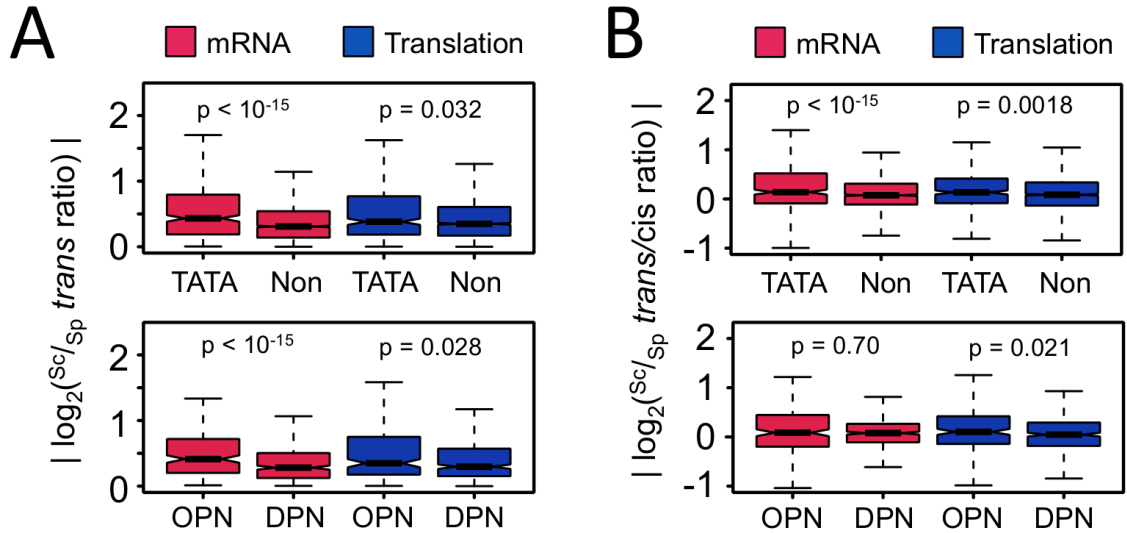

**Supplemental Figure S15.** (A) Orthologs whose promoters contain either TATA boxes (TATA) or occupied proximal nucleosome regions (OPN) show more *trans*-acting divergence only at the mRNA level when compared to non-TATA promoters (Non) or depleted proximal nucleosome regions (DPNs), respectively. *p* values of the Kruskal-Wallis rank sum test are indicated above each fraction. The marginal significance of the translational level comparisons are no longer significant after correction for multiple tests. (B) Comaparison of the relative *trans/cis* ratio shows a stronger effect of TATA boxes but not OPN on *trans* divergence at the mRNA level as has been previously observed in these hybrids (Tirosh et al. 2009). This pattern is also seen more weakly at the translational level, but could reflect the biases in absolute *trans* ratio due to the large number of measurements required (see Supplemental Fig. S15). Analysis was performed exactly as in Fig. 2C, with the exception that only orthologs analyzed in the parental comparisons were used (see Methods in the main text).

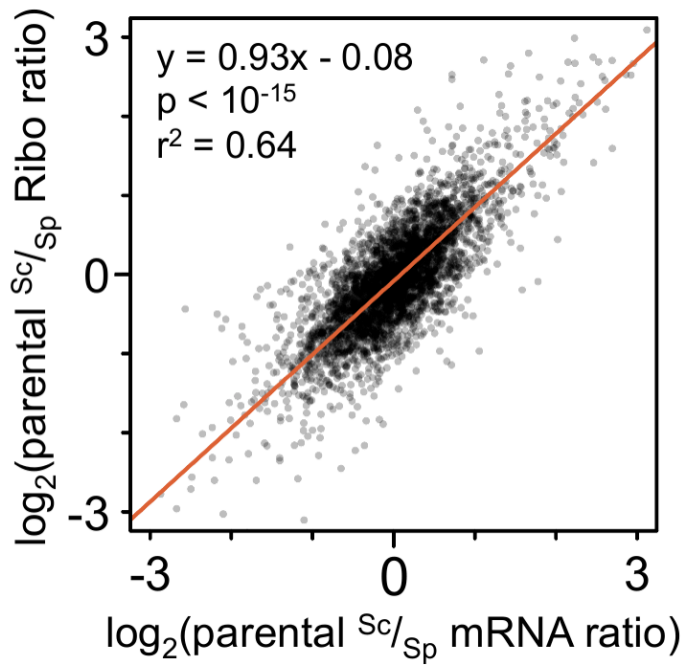

**Supplemental Figure S16.** Opposing divergence across regulatory levels is also observed in the parental samples. The red line indicates the best fit of a linear regression, with equation, p, and $r^2$ values indicated above. The slope is significantly lower than one (95% confidence interval ±0.016), indicating that interspecific ortholog $^{Sc}/_{Sp}$ mRNA ratios tend to overestimate the degree of difference by ~8% relative to that of the Ribo fraction. The higher degree of overestimation observed in the hybrids may reflect the buffering effect of opposing *cis*/*trans* divergence captured in the parental comparison.

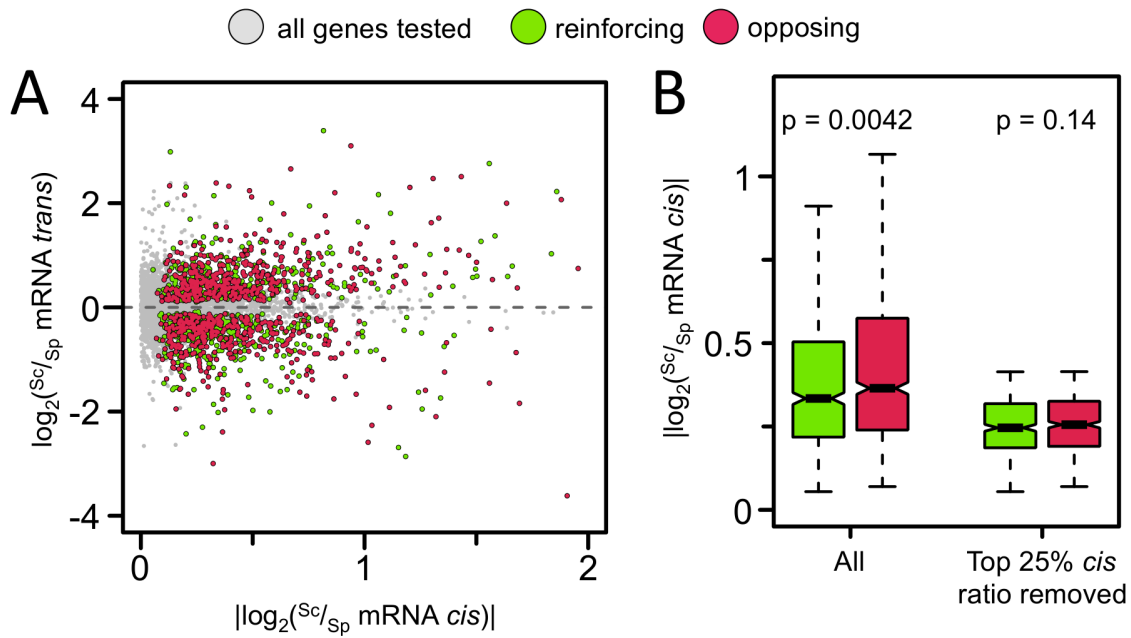

**Supplemental Fig S17.** (A) Scatterplot of $|\log_2(^{Sc}/_{Sp}$ mRNA *cis*$)|$ vs. $\log_2(^{Sc}/_{Sp}$ mRNA *trans*$)$ of orthologs with reinforcing or opposing *cis*/*trans* divergence at the mRNA level. (B) We observed a slight, but significantly higher absolute $^{Sc}/_{Sp}$ mRNA *cis* ratio among orthologs with opposing divergence (Kruskal-Wallis rank sum test, p = 0.0042), which may be biological, but is also consistent with a systematic overestimation of *cis* ratios or underestimation of *trans* ratios among orthologs with strong ASE. However, removal of the top 25% absolute $^{Sc}/_{Sp}$ mRNA *cis* ratio orthologs from the analysis removes this effect, and yet we still observed a significant excess of opposing divergence (406 reinforcing vs. 500 opposing, $\chi^2$ = 9.8, p = 0.0018), which are the values presented in the main Results section. The reciprocal comparison of the distributions of $|\log_2(^{Sc}/_{Sp}$ mRNA *trans*$)|$ in reinforcing vs. opposing orthologs indicates that they are not significantly different from one another when either comparing all ratios (p = 0.065) or with the top 25% of ratios removed (p = 0.61; not shown).

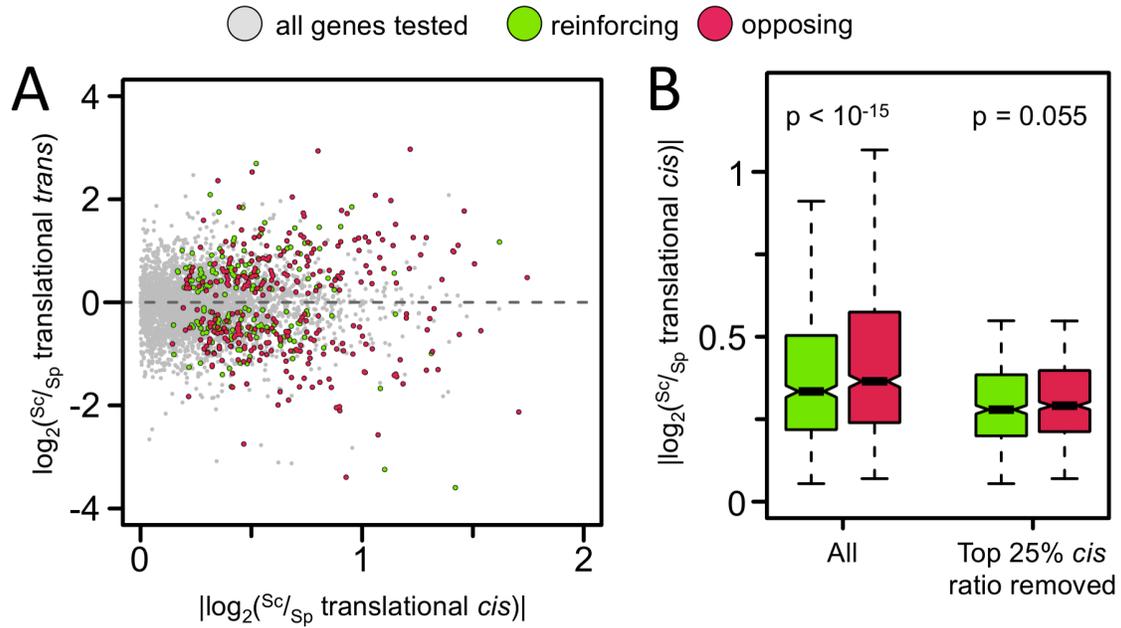

**Supplemental Figure S18.** (A) Scatterplot of $|\log_2(^{Sc}/_{Sp}$ translational *cis*$)|$ vs. $\log_2(^{Sc}/_{Sp}$ translational *trans*) of orthologs with reinforcing or opposing *cis*/*trans* divergence at the translational level. (B) At this level, we observed a strong relationship of higher absolute $^{Sc}/_{Sp}$ translational *cis* ratio among orthologs with opposing divergence (Kruskal-Wallis rank sum test, $p < 10^{-15}$), which could reflect the amount of variability that is included in the estimate of the translational $^{Sc}/_{Sp}$ *trans* ratio (i.e., parental $^{Sc}/_{Sp}$ Ribo - (hybrid $^{Sc}/_{Sp}$ Ribo + parental $^{Sc}/_{Sp}$ mRNA). Again, removal of the top 25% absolute $^{Sc}/_{Sp}$ translational *cis* ratio orthologs from the analysis removes this effect, and there is no evidence of an excess of either reinforcing or opposing divergence (94 reinforcing vs. 91 opposing, $\chi^2 = 0.049$, $p = 0.83$), which are the values presented in the main Results section.